\documentclass[aps,prx,onecolumn,preprint,letterpaper]{revtex4}

\usepackage{appendix}
\usepackage{graphicx}   
\usepackage{import}                         
\usepackage{epstopdf}
\usepackage{amsmath} 
\usepackage{bm}
\usepackage{amssymb}
\usepackage{quotes}
\usepackage{transparent}
\usepackage{dcolumn}
\usepackage{multirow}
\usepackage{cancel} 
\usepackage{mdframed}
\usepackage{color}
\usepackage{bm}
\usepackage{dsfont}
\usepackage{slashed}
\begin{document}
\rmfamily

\title{Complete condensation of photon noise in nonlinear dissipative systems}
\author{Nicholas Rivera$^{1}$, Jamison Sloan$^{2}$, Yannick Salamin$^{2}$, and Marin Solja\v{c}i\'{c}$^{1,2}$}

\affiliation{$^{1}$Department of Physics, MIT, Cambridge, MA 02139, USA.  \\
$^{2}$Research Laboratory of Electronics, MIT, Cambridge, MA 02139, USA. }

\maketitle

\textbf{Fock states are the most fundamental quantum states of bosonic fields, forming an important basis for understanding their quantum dynamics. As energy and number eigenstates, they have an exactly defined number of quanta, and most faithfully express the particle nature of fields. These properties make them attractive for many applications in metrology \cite{davidovich1996sub,thomas2011real}, communication \cite{teich1989squeezed}, and quantum simulation and information processing \cite{aaronson2011computational,lund2014boson,huh2015boson,wang2017high,hamilton2017gaussian, brod2019photonic, wang2020efficient}. Yet, Fock states are notoriously difficult to generate \cite{hofheinz2008generation, wang2008measurement, rempe1990observation, varcoe2000preparing, sayrin2011real}. The problem is especially acute in optics, where it is difficult to deterministically produce Fock states with more than a single photon, let alone at macroscopic scales. This is in part due to a dearth of mechanisms to produce large Fock states, as well as the deleterious effects of linear dissipation. Here, we introduce a detailed theory of a new effect in the physics of nonlinear bosons, arising from the interplay of dissipation and Kerr nonlinearity. In this effect, a nonlinear resonance is dissipationless when it has a particular number of quanta (e.g., photons) inside it, and  lossy otherwise. This loss, which results from nonlinear interference, leads to several new quantum statistical effects. For example, it leads to spontaneous condensation of intensity noise, which may enable generation of large Fock and extremely photon-number-squeezed states of light. We also show how this effect has implications for new classes of optoelectronic devices such as lasers, which can stabilize extremely low-noise states in an equilibrium between gain and the nonlinear loss that we introduce. Throughout the text, we present examples of systems that may realize these effects. In one, we show how the nonlinear dissipation could lead to optical Fock states of $n=1000$, while in another, we show how conventional laser architectures could be used to generate macroscopic light ($>10^{12}$ photons) with nearly 95\% less noise than the standard quantum limit.}

\begin{figure}[t]
    \centering
    \includegraphics[width=0.75\textwidth]{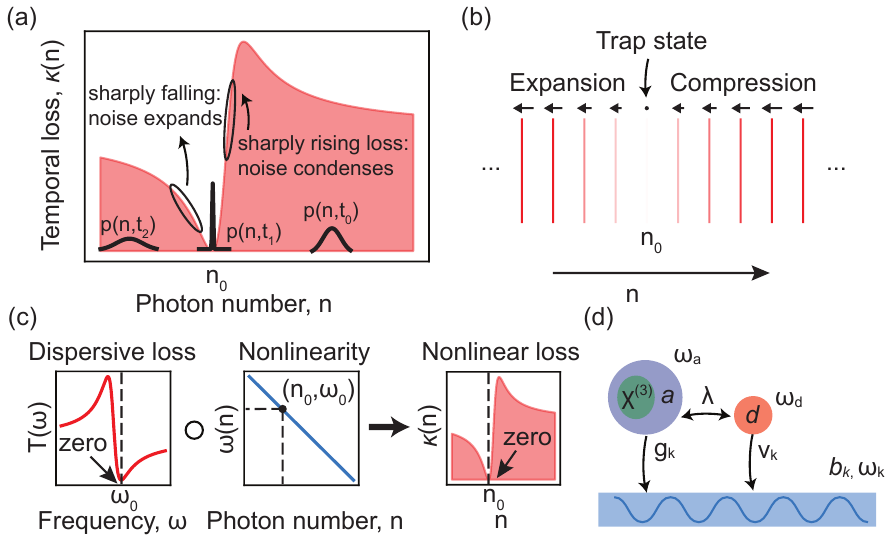}
    \caption{\textbf{Photon noise condensation and Fock state generation in systems with sharply nonlinear loss.} (a) A nonlinear resonance whose loss rate (red) depends on photon number will have its photon number fluctuations compress as it decays, if it falls through a region of sharply rising loss. This is represented by the temporal evolution of the photon probability distribution (black) for different times (with $t_0 < t_1 < t_2$). If the loss has a zero for some photon number $n_0$, the noise condensation is perfect and the system approaches a Fock state of $n_0$ photons. (b) This can be understood through the $n$-dependent rate of transitions from $n$ to $n-1$ photons (arrows denote magnitudes, lines denote states on the Fock ladder). The gradient of the rates (loss ``sharpness'') dictates the magnitude of compression, expansion, or trapping of the distribution.  (c) The requisite nonlinear loss can be understood as arising from a “composition” of a frequency-dependent loss and an intensity-dependent cavity resonance frequency (e.g., due to Kerr nonlinearity). (d) Example of one of the many systems that could realize a loss of the form shown in (a): two resonances coupled to a common continuum, in which one is linear ($d$) and one is nonlinear ($a$). A zero surrounded by a region of sharp loss arises due to destructive (Fano) interference between two leakage pathways for $a$ which can become perfect for a precise number of photons in $a$ (namely, $n_0$).}
    \label{fig:fock1}
\end{figure}

We start by describing the new effects and the intuition behind them. Consider a nonlinear resonance with dissipation (e.g., a leaky mode of a nonlinear cavity). Suppose that the loss rate, $\kappa(n)$ of the mode depends on the number of photons $n$ in the way shown in Fig. 1a. Namely, the nonlinear loss should have regions where the loss increases rapidly with intensity. Ideally, the loss also has a zero for some special photon number $n_0$. In such a system, Fock and highly squeezed quantum states of light can be created. To see how, consider the time evolution of the probability $p(n)$ that the resonance has $n$ photons. If the distribution is concentrated above the minimum of the loss at time $t_0$ (pictured in Fig. 1a), it will eventually fall through the region of sharply increasing loss. This will cause the probability distribution to condense, because the tail of the distribution on the high-number side moves towards lower photon numbers faster than the tail on the low-number side (see Fig. 1b). On the other hand, if the distribution falls through a region of decreasing loss, the distribution will expand (by similar reasoning). If the loss has an exact zero at photon number $n_0$, then Fock states of photon number $n_0$ are created because the probability distribution will get stuck: it cannot move towards lower photon numbers, while the high-number tail gets pushed towards the zero. The special nonlinear loss required to realize the effect can arise by a combination of (1) frequency-dependent loss (for example, if an element of the cavity has frequency-dependent transmission) and (2) Kerr nonlinearity, which leads to a number-dependent resonance frequency (because the index of refraction, and thus the resonance frequency of the cavity depends on the intensity or equivalently the cavity photon number). As we will show rigorously, the frequency-dependent ``loss'' and the number-dependent ``frequency'' compose (in the sense of function composition) to create just the right number-dependent loss (as illustrated schematically in Fig. 1c). For example, for a certain number of photons $n_0$ in the cavity, the resonance frequency $\omega(n)$ will be exactly $\omega_0$, corresponding to the zero of the transmission in Fig. 1c, and thus at $n_0$ photons, the cavity becomes lossless. As we shall discuss later (in Figs. 3 and 4), this nonlinear loss can be then used in place of conventional linear loss in any device that establishes equilibrium between pumping and damping (e.g., a pumped cavity, or a laser). For example, when the nonlinear loss of Fig. 1a is used in place of linear loss in a laser, the equilibrium state of the cavity photons that is established has very low intensity noise (approaching a Fock state).

In what follows, we provide details to the picture painted above. We start by identifying a broad class of physical systems, of the form schematically illustrated in Fig. 1d, that can implement the proposed nonlinear loss and noise condensation effects. We explicitly show, on the basis of a quantum optical theory of nonlinear dissipation, how the described effects arise in this class of systems. Although the effect illustrated in Figs. 1a-c can appear in many more systems than the one shown in Fig. 1d, focusing on the particular type of system shown in Fig. 1d has the benefit of allowing us to rigorously prove the existence of the effect in a way that makes the assumptions and approximations clear. In the main text, we summarize those key results of the theory that underlie the analysis of the examples that we discuss. The Supplementary Information (SI) systematically develops the theory in detail, showing how the effects can be derived from several approaches, all of which are in agreement: master equation methods (SI pgs. 6-13), quantum Langevin methods (SI pgs. 17-21, 39-45), and exact numerical solutions (SI pgs. 27-31). 


\textbf{Theory.} A broad class of physical systems which displays these effects is schematically illustrated in Fig. 1d: one nonlinear oscillator (with annihilation operator $a$), and one linear oscillator (annihilation operator $d$ and frequency $\omega_d$), coupled to a common continuum of bath oscillators (annihilation operators $b_k$ and frequencies $\omega_k$; $k$ indexing the continuum). The couplings of $a$ and $d$ to the continuum are respectively $g_k$ and $v_k$. The Hamiltonian of this general class of systems is:
\begin{equation}
    H/\hbar = \Omega(a^{\dagger}a) + \omega_d d^{\dagger}d + \left(\lambda ad^{\dagger} + \lambda^*a^{\dagger}d\right) + \sum\limits_k \omega_k b^{\dagger}_kb_k +\sum\limits_{k} \left(X_k b_k^{\dagger} + X_k^{\dagger} b_k\right),
\end{equation}
where we have considered the case of an intensity-dependent nonlinear resonance in which the energy of $n$ excitations (photons) is given by $\Omega(n)$. For a Kerr nonlinear system, $\Omega(a^{\dagger}a) = \omega_a(1+\beta + \beta a^{\dagger}a)$, with $\beta\omega_a$ the nonlinear strength of a single photon. The operator $X_k = g_k a + v_k d$ appearing in the Hamiltonian reflects the coupling of $a, d$ to the common continuum. We consider the standard case in which the bath has negligible memory and may be taken to be in the vacuum state. Let us further consider systems for which the response time of the linear resonance ($\gamma = 2\pi \rho_0 v^2$, with $\rho_0$ the density of bath states) is much shorter than that of the nonlinear resonance ($\kappa = 2\pi \rho_0 g^2$): then, $d$ can be adiabatically eliminated, admitting a simple equation for the dynamics of $a$ alone. This is a key approximation in the theory of this system.

A key result is the equation of motion for the reduced density matrix of $a$ (denoted $\rho$) (see SI pgs. 6-13 for derivation):
\begin{equation}
    \dot{\rho} = -\sum\limits_{n=0}^{\infty} n(\mu_n T_{n,n}\rho + \mu^*_n \rho T_{n,n} ) + \sum\limits_{m,n=0}^{\infty} \sqrt{m(n+1)}(\mu_m + \mu_{n+1}^*)T_{m-1,m}\rho T_{n+1,n},
\end{equation}
where $T_{m,n} \equiv |m\rangle\langle n|$, $\mu_n = \frac{1}{2}\kappa - \frac{G_+G_-}{i(\omega_d - \omega_{n,n-1}) + \gamma/2}$, $G_{-} \equiv i\lambda + \frac{1}{2}\sqrt{\kappa\gamma}$, $G_{+} \equiv i\lambda^* + \frac{1}{2}\sqrt{\kappa\gamma}$, and $\omega_{n,n-1} = \Omega(n) - \Omega(n-1)$. While Eq. (2) governs the entire evolution of $a$, we focus here on the the probability $p(n) \equiv \langle n|\rho|n\rangle$ that the nonlinear resonance has $n$ photons. Such probabilities, when more tightly concentrated than the Poisson distribution (so that the variance $(\Delta n)^2 < \bar{n}$, with $\bar{n}$ the mean number of photons), correspond to a sub-Poissonian (number-squeezed) states of light that have no classical analog \cite{mandel1995optical, davidovich1996sub, walls2007quantum} ($\Delta n = 0$ corresponds to a Fock state). The probabilities evolve as: 
\begin{equation}
    \dot{p}(n) = -L(n) p(n) + L(n+1) p(n+1) ,
\end{equation}
where $L(n) \equiv 2n\text{Re }\mu_n$, the rate of transitions from the cavity state with $n$ photons to that with $n-1$ photons, is given by
\begin{equation}
    L(n) = n\left(\frac{\kappa\delta_n^2 + \gamma|\lambda|^2 + 2\sqrt{\kappa\gamma}\delta_n \text{Re }\lambda}{\delta^2_n + \gamma^2/4}\right) \equiv n\kappa(n),
\end{equation}
where $\delta_n = \omega_{n,n-1}-\omega_d$. Eqs. (3) and (4) describe a process of nonlinear dissipation in which excitations decay at a rate $\kappa(n)$ which depends on the number of excitations. 

\begin{figure*}[t]
    \centering
    \includegraphics[width=0.9\textwidth]{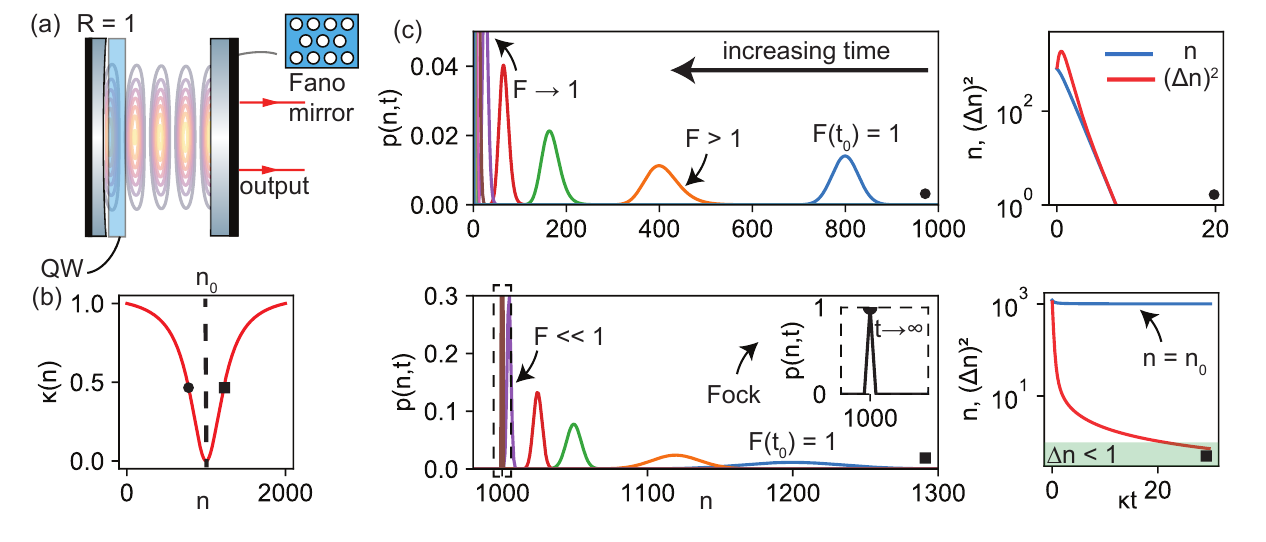}
    \caption{\textbf{Photon noise condensation in systems with sharply nonlinear loss.}  (a) Example system to realize the effect of interest: a nonlinear resonance (e.g., an exciton polariton) coupled to a mirror with an internal resonance (a ``Fano mirror'') with a single loss channel (temporal loss shown in (b)). (c) Time-dependent photon probability distributions for $\bar{n}(0) = 800$ (top) and $\bar{n}(0) = 1200$ (bottom), as well as mean and variance. For $\bar{n}(0) = 800$, the system tends to the vacuum state, while for $\bar{n}(0) = 1200$, the system tends to a Fock state of 1000 photons. In this example: $\beta = 5 \times 10^{-7}, \kappa = 10^{-5}, \gamma = 5 \times 10^{-4}, \omega_d = (1+\delta)$, with $\delta = 10^{-3}$, in units of the lower polariton frequency, 1.47 eV.}
    \label{fig:fock2}
\end{figure*}

The resulting intensity-dependent loss curve $\kappa(n)$ is exactly of the form shown in Fig. 1a. The loss displays a zero for some photon number $n_0$. Away from $n_0$, the loss sharply increases. Much of the behavior of Eq. (4) can be understood from the \emph{linear} equation for the mean values of $a$ and $d$, denoted $\bar{a}$ and $\bar{d}$, which reads (see SI pages 14-15):
\begin{equation}
\begin{pmatrix}
\dot{\bar{a}} \\
\dot{\bar{d}}
\end{pmatrix} = \left[-i\omega_d -\begin{pmatrix}
i\delta + \frac{1}{2}\kappa && i\lambda^* + \frac{1}{2}\sqrt{\kappa\gamma} \\
i\lambda + \frac{1}{2}\sqrt{\kappa\gamma}  && \frac{1}{2}\gamma 
\end{pmatrix} \right]
\begin{pmatrix}
\bar{a} \\
\bar{d}
\end{pmatrix}.
\end{equation}
The dissipation rates of the two coupled modes, for $\kappa \ll \gamma$, are of order $\kappa$ and $\gamma$, as expected. The coupled mode with decay rate $O(\gamma)$ decays very rapidly, and can be ignored. The other mode (which is $a$, to order $\sqrt{\kappa/\gamma}$), has a decay rate which is simply the $\kappa(n)$ of Eq. (4), taking $\delta \rightarrow \delta_n$. The problem of two \emph{linear} resonances coupled to a common continuum, as formulated in Eq. (5), is known to yield vanishing losses for one of the eigenvalues, resulting from interference of two leakage paths for $a$: one in which $a$ passes directly to the continuum, and one in which $a$ transits through $d$ before going to the continuum. This is connected to effects of appreciable recent interest in photonics, such as the Fano effect \cite{fan2002analysis,fan2003temporal,limonov2017fano} and bound states in the continuum \cite{hsu2013observation,hsu2016bound}. Further support for this connection is provided in SI, pages 13-16. The role of nonlinearity is to bind the leakage amplitudes to the excitation number in $a$, such that: for some ``magic'' number of excitations $n_0$ in $a$, the interference is perfect and $a$ is lossless (vaguely reminiscent of electromagnetically-induced transparency \cite{fleischhauer2005electromagnetically}). Stated quantitatively, in the limit $\kappa \ll \gamma$, the Fano transmission profile and the nonlinear Kerr shift ``compose'' (as illustrated in Fig. 1c), converting a linear loss $-$ which introduces intensity fluctuations $-$ into a nonlinear loss, which is known to allow for the possibility of number squeezing \cite{yariv1990self,kitching1994amplitude,mogilevtsev2013nonlinear,thornton2019coherent}. What will distinguish the nonlinear loss of Fig. 1a from previously explored nonlinear losses (e.g., based on multi-photon absorbers \cite{walls1990amplitude,ritsch1990quantum,wiseman1991noise,ritsch1992quantum}), as well as other nonlinear effects such as squeezing in parametric oscillators \cite{andersen201630, walls1983squeezed, bondurant1984squeezed}, is that the number squeezing can in principle be complete, yielding a Fock state of $n_0$ photons. 

There are many physical systems that can realize the type of loss derived here, yielding many opportunities. For example, the loss of Eq. (4) could be realized in a Kerr-nonlinear cavity formed by one perfectly reflecting mirror and one mirror with a frequency-dependent transmission (as for example in a photonic crystal mirror or an etalon): see SI pgs. 16-17 for further discussion. The source of the loss does not need to be transmission: it can also arise due to internal absorption. Absorbers with more complex frequency-dependent absorption lineshapes (such as from electromagnetically induced transparency) may also display the type of zeros which would be amenable to the effects described here \cite{fleischhauer2005electromagnetically}. Our theory is also readily extendable to the case of more general filters with more complex transmission profiles (e.g., a Bragg mirror): in that case, the nonlinear loss is dictated by the frequency-dependent transmission of that system, evaluated at the nonlinear resonator frequency. 

\begin{figure*}[t]
    \centering
    \includegraphics[width=1\textwidth]{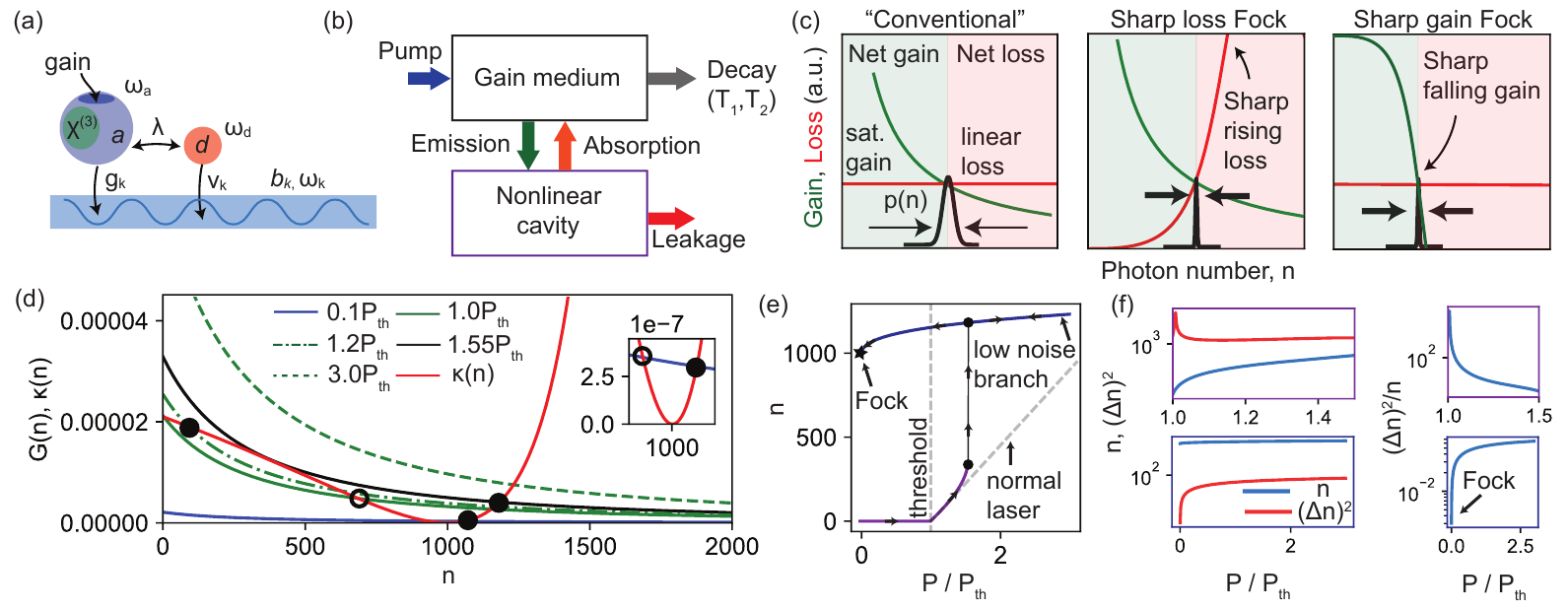}
    \caption{\textbf{The Fock laser.} (a) Components of a general Fock laser, which consists of a pumped gain medium and a nonlinear cavity, interacting via absorption and emission of cavity photons by the gain medium. (b) Energy flows between components of the Fock laser. The cavity leakage is of the sharp form in Fig. 1a. (c, left) Saturable gain and linear loss (corresponding to a conventional laser) leads to Poissonian photon statistics well-above threshold. (c, middle) On the other hand, saturable gain, combined with sharply rising loss, leads to condensation of the photon probability distribution, as in Fig. 1, except now in the steady-state. (c, right) The same condensation also holds when the gain sharply decreases and the loss is linear. (d) Gain and loss curves for a Fock laser for different values of the pump intensity. (e) Mean value of the intracavity photon number as a function of pump strength, relative to threshold. (f) Mean and variance, as well as Fano factor, for the two branches of the input-output curve of (e). Parameters used in this plot are $\beta = 5 \times 10^{-5}, \kappa = 10^{-5}, \gamma = 2 \times 10^{-3}, \omega_d = (1+\delta)$, with $\delta = 0.04$ (in units of the lower polariton frequency $\omega_{\text{LP}}$). Detailed gain and cavity parameters are provided in the SI, pg. 37.}
    \label{fig:fock3}
\end{figure*}

\begin{figure}[t]
    \centering
    \includegraphics[width=0.75\textwidth]{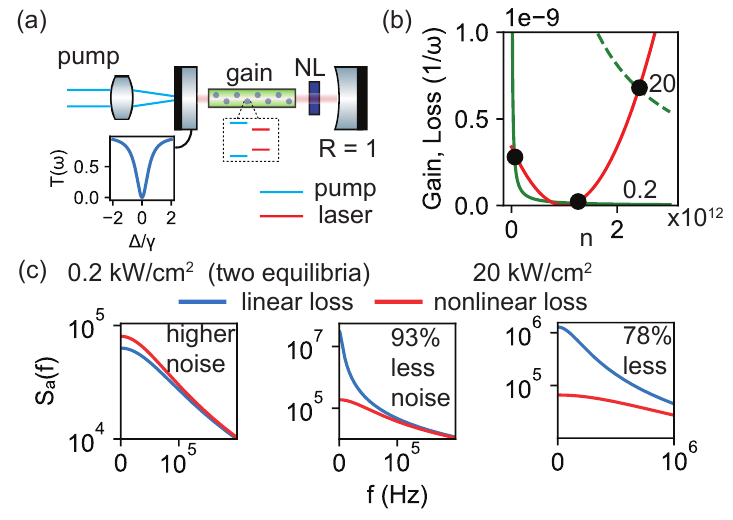}
    \caption{\textbf{Fock lasers in the macroscopic regime and large suppression of photon noise in a common laser architecture.} (a) A macroscopic implementation of a Fock laser based on a diode-pumped solid-state laser with a sharply-varying transmissive element and a nonlinear crystal. (b) Gain-loss diagrams with black circles showing stable equilibria for different pump intensities. (c) Cavity amplitude-noise spectra as a function of frequency for different pump intensities. For intermediate pump intensities, the overall noise reduction can be nearly 95\% of the shot-noise limit with $10^{12}$ photons. The frequency-dependent noise can be reduced by as much as 100-fold for low frequencies. Parameters used in this plot are $\beta = 5 \times 10^{-18}, \kappa = 8 \times 10^{-5}, \gamma = 10^{-2}, \omega_d = (1+\delta)$, with $\delta = -10^{-5}$ (in units of the lasing frequency, 1.17 eV). Detailed gain and cavity parameters are provided in the SI, pg. 37.}
    \label{fig:fl1}
\end{figure}

\textbf{Transient noise condensation.} The unique form of this nonlinear dissipation leads directly to the new quantum statistical effects reported here. One such effect is transient noise condensation. Consider the evolution of the photon probability distribution $p_n(t)$ due to free nonlinear dissipation. We consider initial conditions which are purely diagonal, corresponding to de-phased light, so that the density matrix is specified at all times by the probabilities. For concreteness, consider an initially Poissonian distribution of light (as from e.g., an ideal laser pumped well-above threshold). As per the discussion surrounding Figs. 1a,b, we expect that an initially Poissonian distribution with mean photon number above $n_0$ should rapidly squeeze and approach a Fock state $-$ in stark contrast to the textbook case of linear loss ($\kappa(n) = \kappa$) \footnote{In the case of linear loss, an initially Poissonian photon probability distribution will stay Poissonian, while a Fock state will have its relative fluctuations (measured by its Fano factor, $F = (\Delta n)^2 / \bar{n}$) increase over time. In particular, a Fock state will evolve into a binomial distribution (with success probability $e^{-\kappa t}$, such that $F = 1 - e^{-\kappa t}$ goes to 1 (the Poisson value) as $t \rightarrow \infty$).}. Meanwhile an initial distribution below $n_0$ should expand and eventually become Poissonian.

These intuitions are confirmed by direct solution of Eq. (3) for the photon probabilities. In Fig. 2, we show the time-evolution of the photon probability distribution, as well as the mean and variance, for an example system. The parameters taken are characteristic of systems of exciton-polaritons (arising from strong coupling of a quantum well to a cavity), which have been shown to realize dissipative Kerr Hamiltonians similar to Eq. (1) (without ``$d$'') \cite{fink2018signatures, delteil2019towards}. The strong nonlinearities characteristic of such systems derive from Coulomb interactions between excitons. By coupling the polaritons (representing ``$a$'') to a frequency-dependent loss, the system of Eq. (1) may be realized. This may be done e.g., by coupling the system to a resonator-waveguide system, introducing an absorber, or having one of the cavity mirrors be frequency-dependent (all possible manifestations of ``$d$''). The evolution is shown for two distinct (Poissonian) initial conditions: one in which the mean photon number is below $n_0 = 1000$, and one in which it is above. The case where $\bar{n}(0) < n_0$ does not lead to any noise reduction: after becoming slightly super-Poissonian, the statistics become Poissonian as the amplitude decays to zero. In contrast, when $\bar{n}(0) > n_0$, the variance decays much faster than the mean, ultimately approaching a Fock state of $n_0 = 1000$ photons (Fig. 2c, inset).

It is important to understand that, due to the ``one-way'' nature of loss, residual linear loss, as well as any external effects that cause coupling to lower-photon number states, will destabilize the trapped state and limit the noise condensation. However, even when there is no longer a zero of the loss, heavily sub-Poissonian states can result -- provided that the distribution falls through a region where the loss is sharply increasing. Such states can still fall far below the classical noise limit (beyond number-squeezing experimentally realized thus far), and are still useful for some of the applications described earlier. To explain this, we refer to the equation of motion for the mean and the variance. In the approximation where $\Delta n \ll \bar{n}$:
\begin{align}
    \dot{\bar{n}} &= -L(\bar{n}) \nonumber \\
    \dot{(\Delta n)^2} &= L(\bar{n}) - 2L'(\bar{n})(\Delta n)^2,
    \label{eq:stats_dot}
\end{align}
where $L' \equiv dL/dn$. When $L'(\bar{n}) >  \kappa(\bar{n})$, the variance will decay faster than the mean, and Poissonian light can become sub-Poissonian. This inequality can be achieved by means of a sharply increasing loss (left-hand side of Eq. (6)) and/or a loss coefficient which goes to zero (right-hand side). 

\textbf{Fock lasers.} It is of great interest to stabilize large Fock or sub-Poissonian states in time. This can be achieved by establishing an equilibrium between a pump of energy (e.g., gain) and the nonlinear dissipation. This line of thinking motivates us to introduce and analyze the ``Fock laser,'' shown in Fig. 3a, b: it consists of a pumped gain medium with feedback from an optical resonator. Unlike a conventional laser, the Fock laser uses a resonator with the dissipation of Eq. (4). This loss introduces a new saturation mechanism for the laser which fundamentally differs from that provided by saturable gain. In particular, we will show conditions under which the new saturation leads to steady states with far lower noise (even approaching Fock states) than would be expected (e.g., from an ideal conventional laser, with Poissonian fluctuations of cavity photon number). The number-squeezing can be quite extreme, with examples in the main text displaying nearly 15 dB squeezing over \emph{all} frequencies and over 20 dB squeezing at low frequencies. In the SI (Fig. S4), we show how a laser with this nonlinear loss could present over 30 dB all-frequency squeezing. The theory of lasers with nonlinear loss of the type introduced here is developed in the SI, pgs. 21-27.

The operating principle of the Fock laser is illustrated in Fig. 3c, where we plot the gain and loss coefficients as a function of cavity photon number for a conventional laser (with linear dissipation) versus a Fock laser (with nonlinear dissipation). We consider the ideal case of a single-mode laser in which technical noise due to pump, mechanical, and thermal fluctuations is negligible (due to e.g., active stabilization \cite{black2001introduction}). The mean photon number $\bar{n}$ in the cavity corresponds to where gain balances loss. Thus, the photon probability distribution will be centered around $\bar{n}$. The fluctuations will differ in the two cases, even when the magnitude of the gain and loss (at $\bar{n}$) are identical. The fluctuations are related to the angle of intersection between the gain and loss curves. If the curves intersect steeply, then a small change in photon number leads to a large differential between gain and loss (in absolute value). It is expected that the laser will not occupy such states with high probability, preferring states in equilibrium between gain and loss. Thus, a sharply increasing loss (and/or sharply decreasing gain; illustrated in Fig. 3c but not explored further) leads to suppression of fluctuations beyond those of the conventional laser. This pictorial intuition becomes quantitative in the case of a gain medium where the inversion decay time is fast compared to cavity losses. Then, the equilibrium photon probability distribution is approximately $p(n) \approx  e^{-\frac{(n-\bar{n})^2}{2(\Delta n)^2}}/\sqrt{2\pi (\Delta n)^2}$, where (see SI pgs. 24-26):
\begin{equation}
    \Delta n = \frac{1}{\sqrt{-\frac{d}{dn}\frac{G(n)}{\kappa(n)}\Big|_{\bar{n}}}}  \approx \frac{1}{\sqrt{\kappa'(\bar{n})/\kappa(\bar{n}})}.
\end{equation}
Here, $G(n)$ is the temporal gain coefficient, and the approximation holds when the loss varies much more sharply than the gain. This equation confirms that: if $\kappa$ sharply increases relative to its equilibrium value, the photon noise will be suppressed, and thus, the loss introduced in Fig. 1b facilitates the generation of low-loss equilibrium states. It is interesting to point out that the condition for non-classicality $F = (\Delta n)^2/\bar{n} < 1$, corresponds to $\kappa(\bar{n})/(\bar{n}\kappa'(\bar{n})) \approx \kappa(\bar{n})/L'(\bar{n}) < 1 \implies L'(\bar{n}) >  \kappa(\bar{n})$, which was precisely the condition for transient noise condensation (see Eq. (6)). An important corollary of Eq. (7) is that in order to have $\Delta n \sim 1$, one requires the loss coefficient to change by an amount comparable to itself, over a variation of one photon.

An example of the output characteristics of a Fock laser is shown in Figs. 3(d-f), for a nonlinear resonator similar to that of Fig. 2, integrated with a gain medium (for concreteness, parameters describing the gain are those characteristic of molecular dyes). Much can be understood from the gain-loss curves, plotted in Fig. 3d, where the gain coefficient is shown for different pump intensities. Crudely speaking, stable equilibria exist at values $\bar{n}$ where $G(\bar{n}) = \kappa(\bar{n})$ and $G(\bar{n}^+) < \kappa(\bar{n}^+)$. Characteristic of these non-monotonic loss profiles are (1) multiple stable equilibria (here, at most two) and (2) stable lasing equilibria with finite photon number even when the pump is below threshold (in other words, when $G(0) < \kappa(0)$, so that the system, started from vacuum, cannot have its photon number increase). These multiple equilibria lead to distinct input-output relations between the pump and the steady-state photon number. Above a threshold pump strength $P_{\text{th}}$, the mean photon number increases linearly with pump strength, and the noise is substantially higher than the Poisson level, as expected for a laser weakly above threshold. At a certain intensity (here, about $1.55P_{\text{th}}$), the system discontinuously jumps to a new steady state with much larger photon number, as well as very low noise (about 95\% lower than the standard quantum limit expected from an ideal laser). If we start from this ``low noise branch'' and then lower the pump intensity, the system will follow the purple curve in Fig. 3e, and, as the pump goes down to zero, the stable equilibrium approaches the zero-loss point (see inset of Fig. 3d). This point, in a similar manner to Fig. 2, is accompanied by a low noise equilibrium state, which tends to a Fock state as the zero of the loss is approached. For example, for a pump strength of $0.01P_{\text{th}}$, the noise is 20 dB below the shot noise level, and the photon number uncertainty is roughly 3. 

The Fock laser principle can also be fruitfully extended to truly macroscopic regimes, e.g., ``conventional laser architectures'' employing bulk nonlinearities to generate highly intense sub-Poissonian light. In this case, because the single-photon nonlinear shifts are quite small, one would not be able to generate a state with $\Delta n = O(1)$. However, it is in principle still possible to reduce the noise by a large fraction compared to the standard quantum limit, which, if observed, would yield record squeezing both at a single noise frequency ($>$ 20 dB) and integrated over all frequencies ($>$ 10 dB). An example is presented in Fig. 4 (here, parameters are characteristic of a rare-earth gain medium such as Nd:YAG). Evaluating noise spectra for the cavity photon number for two different intensities, we see that the photon noise (integrated over all frequencies) can drop nearly 95\% below the shot noise limit, but at photon numbers of $10^{12}$, which are clearly macroscopic. Such effects follow directly from Eq. (7), being assisted by a loss which is both sharp and small in magnitude (it is comparable to that offered by state-of-the-art supermirrors \cite{ueda1996ultra}.) 

\textbf{Discussion.}  We have shown that a suitably designed nonlinear dissipation leads to the deterministic generation of macroscopic quantum states of light, such as Fock and heavily photon-number-squeezed (sub-Poissonian) states. The key to deterministically generating such states is a loss which sharply increases away from a minimum (of ideally zero loss). This type of nonlinear dissipation is effectively non-perturbative in intensity, in the sense that it cannot be represented as a low-order Taylor expansion about zero intensity (in the way that the loss associated with two- or few-photon absorption can be). Although such a non-perturbative dissipation is not naturally realized in absorbing materials, we have shown rigorously in a specific class of systems (with the Hamiltonian of Eq. (1)) that the desired nonlinear loss can be constructed by combining frequency-dependent losses with Kerr nonlinearities. This effectively converts a linear loss, which spoils photon-number squeezing, into a nonlinear loss, which can induce it (through the mechanism illustrated in Figs. 1a-c). This composition of frequency-dependent loss and nonlinearity, illustrated in Fig. 1c, suggests a recipe for mitigating the effect of loss in existing experiments: the prescription is to take the dominant loss, and ``make it nonlinear''. The physics described by Eq. (1), namely: dissipative coupling between a linear and a nonlinear resonance, can be realized in a large class of systems both in optics and beyond $-$ implying a great variety of systems to which the physics introduced here can apply.

In optics, the realization of the nonlinear loss of Fig. 1 is perhaps especially addressable now, given recent advances in nanophotonics focusing on the engineering of radiative loss (including dissipative coupling of resonances \cite{leefmans2021topological}). For example, recent work on Fano states and bound states in the continuum (summarized e.g., in \cite{hsu2016bound,limonov2017fano}), when combined with Kerr nonlinearities, may enable realization of the Hamiltonian of Eq. (1) and the loss of Fig. 1b. Nanophotonic systems more broadly (exploiting coupled cavities based on high-Q ring resonators and microspheres \cite{vahala2003optical}, or photonic crystal cavities \cite{akahane2003high}) should enable the construction of almost arbitrary nonlinear losses. Our proposal is also timely in light of a considerable volume of work on ``non-Hermitian'' optical systems with highly-engineered gain and loss \cite{feng2017non,miri2019exceptional}, even including nonlinearity \cite{xia2021nonlinear}.

Compared to other nonlinear loss effects that have been explored for noise suppression (e.g., multi-photon absorbers \cite{walls1990amplitude,ritsch1990quantum,wiseman1991noise,ritsch1992quantum,lathi1999influence}, amplitude-phase coupling \cite{yariv1990self,kitching1994amplitude}, optical bistability \cite{drummond1980quantum}, soliton squeezing \cite{friberg1996observation,kumar2002quantum}) $-$ and even compared to squeezing from second-order nonlinearities (+ coherent displacement) \cite{bondurant1984squeezed,andersen201630} $-$ the nonlinear loss here is the only one we are aware of that can create Fock states. Even in cases where Fock states are not generated, the squeezing often exceeds the theoretical maximum for all of these cases. Another related approach is that used in so-called lasers with ``quiet pumping'' \cite{yamamoto1986amplitude,machida1986observation,richardson1991squeezed, yamamoto1992photon}, where a low-noise pump current is used to reduce the low-frequency noise of a laser. Such approaches lead to at most 50\% noise reduction in the cavity \cite{walls2007quantum}. As compared to other schemes for optical Fock state generation (such as schemes using photon blockade \cite{birnbaum2005photon}, ``unconventional'' photon blockade \cite{flayac2017unconventional}, or engineered driving terms \cite{lingenfelter2021unconditional}), our proposal addresses macroscopic Fock states.

That said, the ideas presented here may also be employed to create one- and few-photon Fock states, by combining single-photon-scale nonlinearities (e.g., using strong coupling \cite{torma2014strong} or Rydberg atoms \cite{peyronel2012quantum}) with Fano interference in a way such that such that $n_0$ (of Fig. 1a-c) is of order one. Such results would represent an exciting milestone in quantum nonlinear optics. All of the present work on creating single-photon scale nonlinearities is also more generally useful in realizing large Fock and sub-Poissonian states, as strong changes in loss over the scale of one photon are needed to get to the ultimate Fock-state limit (as per Eq. (7)). Our scheme can also be applied in the microwave regime, especially in superconducting qubit systems, where reservoirs can be engineered with greater facility  and nonlinearities are quite strong \cite{krantz2019quantum,kockum2019ultrastrong,forn2019ultrastrong}.

Let us also summarize the experimental state of the art: in optics, only one-photon Fock states can be deterministically created thus far \cite{lounis2005single} (using e.g., heralded parametric down conversion, photon blockade, or quantum emitters). Fock states can also be non-deterministically generated by collapsing the wavefunction in the number basis \cite{waks2004direct, cooper2013experimental}. At microwave frequencies, more is possible, and approximate Fock states of around 15 photons have been generated. This is done using cavity quantum electrodynamical interactions with superconducting qubits (essentially adding photons to a cavity ``one at a time'' until linear loss sets in \cite{hofheinz2008generation, wang2008measurement}). Other schemes applied at microwave frequencies include the ``micromaser'' \cite{scully1999quantum, rempe1990observation,varcoe2000preparing} and quantum feedback protocols \cite{sayrin2011real}.

Regarding experimental realization of the effects in optics on macroscopic scales, the systems we discussed (especially Fock lasers) entail a huge design space (see the table on SI pg. 37 to get a sense). It is almost certain that there are better platforms than ones discussed here to realize the physics proposed in this work. The systems we chose were mainly taken for the sake of illustration: to show what would be needed. For the gain medium, an obvious choice to consider is semiconductor gain media which need not be laser pumped, and can provide rather high gain over a very broad frequency range, enabling compatibility with many different nonlinear materials. Another important advantage of semiconductor gain media is that they could be integrated into nanophotonic platforms which present high nonlinearities ($\beta$ at least 10 orders of magnitude higher than the bulk realizations presented, due to the reduced mode volumes \cite{soljavcic2004enhancement}).

Given the generality of the effects introduced here, we expect that the theoretical and experimental development of physical platforms to realize them will provide a great deal of exciting new areas for discovery.

\textbf{Acknowledgements.} The authors acknowledge discussions with I. Kaminer, C. Roques-Carmes, and H. Zhou. This material is based upon work supported in part by the Air Force Office of Scientific Research under the award number FA9550-20-1-0115; the work is also supported in part by the U. S. Army Research Office through the Institute for Soldier Nanotechnologies at MIT, under Collaborative Agreement Number W911NF-18-2-0048. N.R. was supported by fellowships from the Department of Energy (DE-FG02-97ER25308) and the MIT School of Science. J.S. was supported in part by the Department of Defense NDSEG fellowship no. F-1730184536. Y.S. acknowledges support from the Swiss National Science Foundation (SNSF) through the Early Postdoc Mobility Fellowship No. P2EZP2-188091.


\clearpage

\renewcommand{\thepage}{S\arabic{page}} 
\renewcommand{\thesection}{S\arabic{section}}  
\renewcommand{\thesubsection}{S\arabic{section}}  
\renewcommand{\thetable}{S\arabic{table}}  
\renewcommand{\thefigure}{S\arabic{figure}}
\renewcommand{\theequation}{S\arabic{equation}}

\begin{center}
\textbf{\large Supplementary Information for: \\
Complete condensation of photon noise in nonlinear dissipative systems}
\end{center}
\setcounter{page}{1}
\setcounter{equation}{0}
\setcounter{figure}{0}
\setcounter{tocdepth}{1}
\setcounter{secnumdepth}{1}

\tableofcontents

\section{Introduction}

In this Supplementary Information (SI), we develop the theory of nonlinear dissipation and amplification in systems with sharp loss (as well as gain). In contrast to the main text, which summarizes the key theoretical results and focuses primarily on applications of the theory, the SI is meant to provide a detailed account of the theory, providing underlying assumptions, as well as derivations. 

In the section "Quantum theory of a nonlinear resonator with frequency-dependent loss" (pgs. 4-21), we will introduce a nonlinear open system model (and its Hamiltonian, Eq. (1) of the main text) that realizes the sharp loss described in the main text. Then, we develop a master equation (Eq. (2) of main text) to describe dissipation in such systems, showing that it coincides with the type of nonlinear dissipation quoted in the main text (Eqs. (3-5) of main text). From there, we move to derive results related to the statistical dynamics (Eq. (6) of main text)). To close Section I, we develop a quantum Langevin theory of nonlinear dissipation in these systems. The quantum Langevin theory is in correspondence with the density matrix theory, and makes the same predictions as far as the results of the main text are concerned. However, the quantum Langevin approach provides the most convenient starting point for describing fluctuations of lasers. In the Appendix (pgs. 39-45), we develop an independent derivation of the nonlinear loss developed in this work through the Heisenberg equations for the projection operators of a nonlinear resonator. 

In the section "Lasers based on sharply nonlinear loss" (pgs. 21-27), we develop the quantum theory of lasers with nonlinear loss. We derive a set of ``quantum rate equations'' $-$ operator-valued rate equations with fluctuating driving terms (Langevin forces) $-$ to describe inversion and photon number fluctuations of lasers. We then derive amplitude noise spectra describing the photon number fluctuations of the laser cavity to lowest nontrivial order in the mean-field approximation. The treatment provided allows one to account for quantum fluctuations in systems with a wide variety of gain media, including gases, molecular dyes, rare-earth dopants (as in solid-state lasers), and semiconductors. 

In the section "Numerical evidence for the effects predicted in the manuscript" (pgs. 27-31), we provide numerical validation of the analytical theory developed here. In the first part, we show that the Fock- and sub-Poissonian state-generation effects follow from explicit time-evolution of the master equation corresponding to the Hamiltonian of Eq. (1) of the main text (under a white-noise approximation for the reservoir).  In the second part, we provide numerical evidence for the Fock lasing effect. In particular, we show that by modifying the Hamiltonian to include a pumped two-level atom (representing a gain medium), we can create a system that supports steady states (of the Liouvillian) corresponding to low-noise states of light. 

In the section "Summary of main results" (pgs. 31-34), we summarize the main new theoretical results developed in this work, for ease of quotation. In the section "Potential extensions of the theory" (pgs. 34-36), we provide a non-exhaustive list of potential extensions of the work presented here which we believe to be exciting directions of future work. We expect the results derived in the SI to have wider applicability than the Fock- and sub-Poissonian proposals considered in the main text. We believe in particular that the theoretical results concerning the master equations for these dissipative nonlinear systems, as well as the Langevin equations we derive, should provide a useful basis for application to the theory of many more complex optoelectronic device configurations.  Finally, in the section "Supplementary figures" (pgs. 36-38), we provide additional data, as well as detailed lists of parameters for Figs. 3, 4 of the main text. 



\section{Quantum theory of a nonlinear resonator with frequency-dependent loss}

\subsection{Model and Hamiltonian of a system with nonlinear loss}

The starting point in our analysis of loss in a nonlinear resonator with frequency-dependent loss is the specification of the Hamiltonian, which describes the nonlinear cavity, the frequency-dependent end-mirror, and all reservoirs responsible for dissipation of the photon. Let us describe each term in the total Hamiltonian in steps.

\emph{Nonlinear cavity.} We start by describing the cavity. We will assume in all cases that we are under conditions of single-mode lasing, and can thus consider the electromagnetic field of the cavity as described by a single high-$Q$ resonant mode. In the absence of photon nonlinearity, the Hamiltonian of the cavity would be simply $\hbar\omega a^{\dagger}a$, with $\hbar$ the reduced Planck constant, $\omega$ the frequency of the resonant mode, and $a$ ($a^{\dagger}$) the annihilation (creation) operator of the cavity mode. Let us consider now what happens when a nonlinear element is introduced into the cavity. 

Consider for example the case of a nonlinear crystal embedded in the cavity, leading to Kerr nonlinear shifts of the cavity frequency. The resulting cavity Hamiltonian can be written in the form $H_{\text{Kerr}} = \hbar\omega a^{\dagger}a +  \frac{1}{6}\beta\hbar\omega:(a-a^{\dagger})^4:$ \cite{drummond1980quantum}, where $\beta$ is a nonlinear coupling constant, and $::$ denotes normal ordering. In the rotating-wave approximation (i.e., ignoring terms with unbalanced numbers of creation and annihilation operators), the Kerr nonlinearity takes the more commonly stated form $H_{\text{Kerr}} = \hbar\omega\left((1+\beta) a^{\dagger}a + \beta(a^{\dagger}a)^2\right)$ \cite{drummond1980quantum,walls2007quantum}. The cavity eigenstates are Fock states of $n$ photons with energy $E_n \equiv \hbar\omega_n = \hbar\omega\left[(1+\beta)n + \beta n^2\right]$.
The Hamiltonian, in the number basis, may alternatively be written as
\begin{equation}
    H_{\text{Kerr}} = \sum\limits_{n=0}^{\infty} E_n T_{n,n},
\end{equation}
with $T$ a projection operator (projector), which is generally defined as: $T_{i,j} \equiv |i\rangle\langle j|$. We have re-written the Hamiltonian in terms of projectors, as they will play an essential role in our theory of nonlinear lasers. Before moving on to the theory of nonlinear lasers, we point out that in this Kerr resonator, the excitation energy from a state with $n-1$ photons, to a state with $n$ photons, is $\omega_{n,n-1} = \omega(1+2\beta n)$. This is equivalent to the statement in classical nonlinear optics that the frequency of a nonlinear cavity shifts by an amount proportional to the intensity \cite{haus1984waves}. The interaction constant $\beta$ is governed by the overlap integral between the (normalized) cavity mode $\mathbf{u}(\mathbf{r})$ and the third-order nonlinear susceptibility $\chi^{(3)}(\mathbf{r})$  (taken as a scalar here for simplicity). In particular $\beta = \left( \frac{3\hbar\omega}{8\epsilon_0}\right)\int d^3r ~ \chi^{(3)}(\mathbf{r})|\mathbf{u}(\mathbf{r})|^4$. Its characteristic magnitude, for a crystal which fills the cavity, is $\frac{3\hbar\omega}{8\epsilon_0 V}\chi^{(3)}(\mathbf{r})$, with $V$ the mode volume. Before moving on to discuss the other terms in the Hamiltonian, we note that a general \emph{intensity-sensitive} nonlinear cavity will have a Hamiltonian of the form of Eq. (S1) with the appropriate photon-number-dependent energies, and so our treatment applies more generally than to the case of Kerr nonlinearities.

\emph{Cavity losses.} Now we move to a discussion of the terms in the Hamiltonian responsible for the losses of the cavity. For the photon, the reservoirs depend on the exact configuration. In the simplest (and most standard case) the photon is coupled to a single reservoir of far-field modes which convert the cavity photon into the emitted beam. To get the Fock and sub-Poissonian state-generation effect, we must go beyond this single cavity-reservoir coupling. The simplest modification that ``does the job'' is to introduce two resonances ($a, d$, as in Fig. 1 of the main text) that are coupled to the same reservoir. This mutual coupling to the same reservoir allows for the Fano-type interferences well-known from classical optics. This approach was recently used to describe the quantum optics of Fano mirrors in \cite{vcernotik2019cavity} (without nonlinearity). Compared to prior work, we consider the case where one of the resonances is nonlinear. In such cases, Fock-state generation is supported under appropriate conditions.


We now set up the Hamiltonian of the ``nonlinear Fano resonance.'' Let us consider a situation in which one mode (labeled by its annihilation operator $a$, with anharmonic Hamiltonian $H_a$) is coupled to a second mode (e.g., a Fabry-Perot type mode, or a photonic crystal resonance), of frequency $\omega_d$ (labeled by annihilation operator $d$). In many cases, this second resonance $d$ can be thought of as the resonance of an end-mirror of the cavity, and we will occasionally refer to $d$ as the mirror. We take the $d$-resonance to be linear, with Hamiltonian $H_{\text{d}} = \hbar\omega_d d^{\dagger}d$. The two modes in general are coupled by a (beam-splitter) interaction $\hbar(\lambda ad^{\dagger} + \lambda^* a^{\dagger}d)$. Both $a$ and $d$ are also coupled to the continuum of far-field modes $b_k$ outside of the cavity, where $k$ enumerates the continuum of outside modes. For simplicity, we will consider a one-sided cavity, with one wall perfectly reflecting, and one partially reflecting, such that there is only a single input and output ``port.'' Taking $g_k$ and $v_k$ to respectively be the coupling of $b_k$ to $a$ and $d$ , the system-reservoir coupling Hamiltonian may be written as: $H_{\text{res}} = \sum\limits_k \hbar g_k(ab_k^{\dagger} + a^{\dagger}b_k) + \sum\limits_k \hbar v_k(db_k^{\dagger} + d^{\dagger}b_k)$. The total Hamiltonian of the system and reservoir may thus be expressed as:
\begin{equation}
	H/\hbar = H_a + \omega_d d^{\dagger}d + (\lambda ad^{\dagger} + \lambda^* a^{\dagger}d) + \sum\limits_k \omega_k b^{\dagger}_k b_k + \sum\limits_k (g_kab_k^{\dagger} + g_k^*a^{\dagger}b_k) + \sum\limits_k  (v_kdb_k^{\dagger} + v^*_kd^{\dagger}b_k),
\end{equation}
which coincides with Eq. (1) of the main text (defining $H_a = \hbar\Omega(a^{\dagger}a)$ and $X_k = g_k a + v_k d$). The simpler case of a Fabry-Perot mirror (with a symmetric transmission spectrum) is obtained in the limit where the ``direct'' coupling of the cavity mode to the far-field can be neglected ($g_k = 0$), so that the cavity must couple through the mirror if it is to escape into the far-field. The other important standard case is that in which the partially reflecting mirror has a frequency independent reflectivity, which corresponds to the case in which the $d$ cavity has a very fast decay. We note that while the parameters $\lambda, g_k, v_k$ could be in principle be calculated, it is typically impractical to do so, and they may in practice be found by comparing the transmission of the cavity to what is expected from a classical treatment of the cavity transmission (e.g., from temporal coupled mode theory). 

\subsection{Master equation of the nonlinear Fano resonance}

In this section, we derive a master equation to describe the damping of a nonlinear resonator ($a$) due to radiative leakage from a frequency-dependent mirror. The overall Hamiltonian of the system+reservoir ($a$ + $d$ + reservoir) is given by Eq. (S2). To simplify notation, we will define
\begin{equation}
    H_{ad} \equiv H_a + \omega_d d^{\dagger}d + \left(\lambda ad^{\dagger} + \lambda^* a^{\dagger}d\right).
\end{equation}
Let us now derive an equation of motion for the reduced density matrix of $a$ and $d$ (e.g., tracing out the reservoir). To do so, we define the interaction picture operators $\rho_I = e^{iH_{0}t}\rho e^{-iH_{0}t}$ and $V_I = e^{iH_{0}t}V e^{-iH_{0}t}$, with $H_0 = H_{ad} + \sum\limits_k \omega_k b^{\dagger}_k b_k$ and $V = \sum\limits_k \left(X_k b^{\dagger}_k + X_k^{\dagger} b_k\right)$. Then, the equation of motion for the density matrix becomes $\dot{\rho_I} = -\frac{i}{\hbar}\left[V_I, \rho_I \right]$, admitting the iterative solution:
\begin{equation}
    \dot{\rho_I} = -\frac{i}{\hbar}\left[V_I(t), \rho(0)\right] - \frac{1}{\hbar^2}\int\limits_0^t dt'~ \left[V_I(t),\left[V_I(t'),\rho_I(t')\right]\right],
\end{equation}
with $\rho(0) = \rho_I(0)$ being the initial state of the system and reservoir. As we will primarily be interested in the application of this framework at optical frequencies, we will consider the reservoir to be in its vacuum state (i.e., negligible thermal population). The dynamics of the resonator and end-mirror are obtained by taking the partial trace with respect to the bath ($\dot{\rho}_{ad} \equiv \text{tr}_b\rho$), such that 
\begin{equation}
    \dot{\rho}_{ad,I} = -\frac{i}{\hbar}\text{tr}_b\left([V_I(t), \rho(0)]\right) - \frac{1}{\hbar^2}\int\limits_0^t dt'~ \text{tr}_b\left(\left[V_I(t),\left[V_I(t'),\rho_I(t')\right]\right]\right).
\end{equation}
Upon taking the trace with respect to the bath, the term which is linear in $V_I$ will vanish, and the equation of motion becomes
\begin{equation}
    \dot{\rho}_{ad,I} = -\frac{1}{\hbar^2}\int\limits_0^t dt'~ \text{tr}_b\left(V_I(t)V_I(t')\rho_I(t') + \rho_I(t')V_I(t')V_I(t) - V_I(t)\rho_I(t')V_I(t') - V_I(t')\rho_I(t')V_I(t) \right).
\end{equation}
To proceed, we need further approximations. As the coupling of system and reservoir is weak, and the continuum of radiation modes loses memory over a very short timescale (due to its infinite bandwidth), we make the standard Markov approximation. Namely, that $\rho$ factorizes as $\rho_I(t') = \rho_{ad,I}(t')\rho_b(0)$, with $\rho_b$ being the density matrix of the multimode vacuum reservoir. Moreover, due to the weak coupling of $a$ and $d$ to the reservoir, the system-reservoir couplings can be approximated as frequency-independent (such that $g_k \approx g$ and $v_k \approx v$). It follows that the first term, under these approximations, evaluates to $X_I(t)X_I^{\dagger}(t')\rho_{ad}(t')\sum_k e^{i\omega_k(t-t')} = X_I(t)X_I^{\dagger}(t')\rho_{ad}(t')(2\pi\rho_0\delta(t-t'))$, with $\rho_0$ the density of states of the far-field continuum (which under these approximations is frequency-independent). Performing the time-integration yields $X_I(t)X_I^{\dagger}(t)\rho_{ad}(t)$. The other terms are evaluated in a similar fashion, yielding 
\begin{equation}
    \dot{\rho}_{ad,I} = - 2\pi\rho_0\left( X^{\dagger}_I(t)X_I(t)\rho_{ad,I}(t) + \rho_{ad,I}(t)X^{\dagger}(t)X_I(t) - 2X_I(t)\rho_{ad,I}(t)X^{\dagger}_I(t)\right).
\end{equation}
Going back to the Schrodinger picture, one has the equation of motion for the system ($a$ + $d$):
\begin{equation}
    \dot{\rho} = -i[H_{ad},\rho] - 2\pi\rho_0\left( X^{\dagger}X\rho + \rho X^{\dagger}X - 2X\rho X^{\dagger}\right),
\end{equation}
where we have taken $\rho_{ad} \rightarrow \rho$ for simplicity of notation (the bath will no longer enter the equations). 

Eq. (S8) can be taken as the first-principles master equation for the nonlinear Fano resonance, upon which we will make further approximations to analytically isolate the nonlinear loss presented in the main text (e.g., Eqs. (2-4) of the main text). Note that, as compared to standard master-equation descriptions of lossy systems, Eq. (S8) is of a similar Lindblad form, except that the jump operator $X$ couples the two modes. In Section III, where we present ``exact'' numerical evidence for the Fock- and sub-Poissonian state generation effects, we do so by directly solving Eq. (S8) in time. Now, we move to simplify Eq. (S8) further.

We are mainly interested in the limit in which the $d$ resonance responds instantaneously to changes in the frequency of the cavity mode. In other words, in the limit of $\gamma \equiv 2\pi\rho_0v^2$ being the fastest timescale of the problem (so for example, $\gamma \gg \kappa \equiv 2\pi\rho_0 g^2$). Physically, thinking of $d$ as the end-mirror, it refers to a situation where the mirror responds to the instantaneous frequency of $a$ (to which the mirror can immediately respond due to its large bandwidth). Under this condition, we may adiabatically eliminate $d$ from the master equation of Eq. (S8), getting an equation of motion for $a$ alone. 

The adiabatic elimination proceeds along similar lines to the derivation of Eq. (S8): we must look at the evolution of the cavity density matrix to second-order in the coupling between $a$ and $d$. The procedure to arrive at the equation for $a$ is thus similar in spirit to the procedure leading to Eq. (S5). A major difference in execution arises from the fact that the free dynamics of $d$ include damping (which is ``fast''), and so the interaction-picture transformation must include the effect of damping. Therefore, the Liouvillian to be exponentiated contains a Lindblad term. While interaction picture transformations of Liouvillians with Lindblad terms are a ``basic'' part of density-matrix theory, they are not as commonplace in the literature (\cite{carmichael2009statistical} provides a good account). Thus, we shall provide more of the intermediate manipulations than in other sections of the SI. 

The equation of motion for the density matrix (in the Schrodinger picture) may be written as
\begin{equation}
    \dot{\rho} = (\mathcal{L}_0 + \mathcal{L}_1)\rho,
\end{equation}
where 
\begin{equation}
\mathcal{L}_0 \equiv -i[H_a/\hbar + \omega_d d^{\dagger}d,\cdot]-\gamma(d^{\dagger}d\cdot + \cdot d^{\dagger}d - 2d^{\dagger}\cdot d),
\end{equation}
and
\begin{align}
\mathcal{L}_1 \equiv &-i[\lambda ad^{\dagger} + \lambda^* a^{\dagger}d,\cdot]-\kappa(a^{\dagger}a\cdot + \cdot a^{\dagger}a - 2a^{\dagger}\cdot a) \nonumber \\
-&\sqrt{\kappa\gamma}\left((ad^{\dagger}+a^{\dagger}d)\cdot + \cdot(ad^{\dagger}+a^{\dagger}d) - 2\left(a\cdot d^{\dagger} + d\cdot a^{\dagger}\right) \right).
\end{align}
Here, we have introduced the $\cdot$ notation, which indicates how the Liouvillian acts on an operator. For example, for arbitrary operators $X,\rho$, we have: $(X\cdot)\rho \equiv X\rho$ and $(\cdot X)\rho = \rho X$. Terms of the form $(X\cdot Y)\rho$, for arbitrary $X,Y$ should be understood as $(X\cdot)(\cdot Y)\rho = X\rho Y$. The terms Eq. (S11) may also be regrouped to read as:
\begin{align}
\mathcal{L}_1 = & -\kappa(a^{\dagger}a\cdot + \cdot a^{\dagger}a - 2a^{\dagger}\cdot a) \nonumber \\
& -\left(G_- (ad^{\dagger}\cdot) + G_-^*(\cdot a^{\dagger}d) \right)  -\left(G_+(a^{\dagger}d) + G_+(\cdot ad^{\dagger})\right) +2\sqrt{\kappa\gamma}\left(a\cdot d^{\dagger} + d\cdot a^{\dagger} \right),
\end{align}
with $G_- \equiv i\lambda + \sqrt{\kappa\gamma}$ and $G_+ \equiv i\lambda^* +\sqrt{\kappa\gamma}$. This expression proves more convenient for the manipulations that follow.

We now define the interaction picture density matrix $\rho_I$ as 
\begin{equation}
    \rho = e^{\mathcal{L}_0 t}\rho_I,
\end{equation}
so that 
\begin{equation}
    \dot{\rho}_I = e^{-\mathcal{L}_0 t}\mathcal{L}_1e^{\mathcal{L}_0 t}\rho_I \equiv \mathcal{L}_I(t)\rho_I.
\end{equation}
This equation admits an iterative solution of the form
\begin{equation}
    \dot{\rho}_{a,I} = \text{tr}_d\left[\mathcal{L}_I(t) \rho_I(0) \right] + \int\limits^t dt'~\text{tr}_d\left[\mathcal{L}_I(t)\mathcal{L}_I(t')\rho_I(t') \right].
\end{equation}
This equation is considerably simplified in the limit where $\gamma$ is large: in this case, $d$ acts as a broad continuum for $a$ (in other words, as a reservoir). Moreover, $d$ cannot sustain any build-up of excitations, as they damp immediately (on any timescale related to $a$). It follows that from the perspective of $a$, $d$ acts as a vacuum reservoir $|0 \rangle\langle 0|$, and that the state of the joint system may be written in factorizable form: $\rho_{I}(t) \approx \rho_{a,I}(t) |0 \rangle\langle 0|$. This allows us to write Eq. (S15) in the Born-Markov approximation as
\begin{equation}
       \dot{\rho}_{a,I} = \text{tr}_d\left[\mathcal{L}_I(t) \rho_a(t) |0 \rangle\langle 0|\right] + \int\limits^t dt'~\text{tr}_d\left[\mathcal{L}_I(t)\mathcal{L}_I(t')\rho_a(t)|0 \rangle\langle 0| \right]. 
\end{equation}
Here, we have also made an adiabatic approximation, replacing $\rho_a(t')$ with $\rho_a(t)$, since significant contributions to the integrand only arise when $t'$ is within $\gamma^{-1}$ of $t$. Over this range of times, the density matrix of $d$ does not vary. To proceed, we must now evaluate the interaction picture Liouvillian operators to second-order, and then evaluate the integrals that arise. The following interaction-picture transformations for $d$ are used heavily in what follows (see \cite{carmichael2009statistical}):
\begin{align}
 &(d \cdot )_I(t)   = e^{-i\omega_dt-\gamma t}(d \cdot ) \nonumber \\
 &(\cdot d^{\dagger})_I(t) = [(d \cdot )_I(t)]^{\dagger} = e^{i\omega_dt-\gamma t}(\cdot d^{\dagger}) \nonumber \\
 &(d^{\dagger} \cdot )_I(t)  = e^{i\omega_dt}\left(e^{\gamma t}(d^{\dagger} \cdot ) + (e^{-\gamma t}-e^{\gamma t})(\cdot d^{\dagger}) \right)\nonumber \\
 &(\cdot d)_I(t)  = [(d^{\dagger} \cdot )_I(t)]^{\dagger} = e^{-i\omega_dt}\left(e^{\gamma t}( \cdot d) + (e^{-\gamma t}-e^{\gamma t})(d\cdot) \right) .
\end{align}
Similarly, the interaction picture transformations for $a$ are given as
\begin{align}
  &(a \cdot )_I(t) = [(\cdot a^{\dagger})_I(t)]^{\dagger} = \sum\limits_{n=0}^{\infty} \sqrt{n}e^{-i\omega_{n,n-1} t}(T_{n-1,n} \cdot) \nonumber \\
 &(a^{\dagger} \cdot )_I(t) = [(\cdot a)_I(t)]^{\dagger} = \sum\limits_{n=0}^{\infty} \sqrt{n+1}e^{i\omega_{n+1,n} t}(T_{n+1,n} \cdot),
\end{align}
where we have defined the projector $T_{ij} = |i\rangle\langle j|.$ Note that due to the polychromatic nature of $a$ (being anharmonic), this is the most convenient way to express the interaction picture operator. With these identities established, we now evaluate the first- and second-order terms of Eq. (S12). 

As $d$ is in the vacuum state, no terms in $\mathcal{L}_1$ involving $d$ or $d^{\dagger}$ contribute to the first-order term. Therefore, the first order term is simply $-\kappa(a_I^{\dagger}a_I\cdot + \cdot a_I^{\dagger}a_I - 2a_I^{\dagger}\cdot a_I) \nonumber$, and in the Schrodinger picture, gives the expected term $-\kappa(a^{\dagger}a\cdot + \cdot a^{\dagger}a - 2a^{\dagger}\cdot a) \nonumber$. Now we evaluate the second-order term. To proceed, we note that since $\gamma \gg \kappa$, we may neglect contributions of order greater than $\kappa$. Hence, we may completely ignore the first line of Eq. (S12) for the purposes of evaluating the second-order term. After some algebra, one finds that the second order integrand, under the assumption that $d$ is in the vacuum state, is given by:
\begin{align}
 -|G_-|^2& \text{tr}_d\left[a_I(t)d_I^{\dagger}(t)\rho_a(t) |0 \rangle\langle 0| a_I^{\dagger}(t')d_I(t')\right] -G_+G_- \text{tr}_d\left[a^{\dagger}_I(t)d_I(t) a_I(t')d_I^{\dagger}(t')\rho_a(t) |0 \rangle\langle 0|\right] \nonumber \\
 -|G_-|^2& \text{tr}_d\left[a_I(t')d_I^{\dagger}(t')\rho_a(t) |0 \rangle\langle 0| a_I^{\dagger}(t)d_I(t)\right] -(G_+G_-)^* \text{tr}_d\left[\rho_a(t) |0 \rangle\langle 0|a^{\dagger}_I(t')d_I(t') a_I(t)d_I^{\dagger}(t)\right] \nonumber \\
 +2\sqrt{\kappa\gamma}& \text{tr}_d\left[a_I(t)\rho_a(t) |0 \rangle\langle 0|a_I^{\dagger}(t') d_I(t') d_I^{\dagger}(t)\right]  +2\sqrt{\kappa\gamma} \text{tr}_d\left[d_I(t)a_I(t')d_I^{\dagger}(t')\rho_a(t) |0 \rangle\langle 0| a_I^{\dagger}(t)\right].
\end{align}
Plugging in the interaction picture operators of Eqs. (S17) and (S18), and evaluating the $t'$-integral, one arrives at the following final result (in the Schrodinger picture):
\begin{align}
    \dot{\rho} &= -\kappa(a^{\dagger}a\rho + \rho a^{\dagger}a - 2a\rho a^{\dagger}) \nonumber \\
    &+ \sum\limits_{n=0}^{\infty} \frac{n G_+G_-}{i(\omega_d - \omega_{n,n-1})+\gamma}T_{n,n}\rho + \sum\limits_{n=0}^{\infty} \frac{n(G_+G_-)^*}{-i(\omega_d - \omega_{n,n-1})+\gamma}\rho T_{n,n} \nonumber \\
    &- \sum\limits_{m,n=0}^{\infty} \frac{\sqrt{m(n+1)}(G_+G_-)^*}{-i(\omega_d - \omega_{n+1,n})+\gamma}T_{m-1,m}\rho T_{n+1,n} - \sum\limits_{m,n=0}^{\infty} \frac{\sqrt{m(n+1)}(G_+G_-)}{i(\omega_d - \omega_{m,m-1})+\gamma}T_{m-1,m}\rho T_{n+1,n}.
\end{align}
Here, we have taken $\rho_a \rightarrow \rho$, as no further reference will be made to the density operator of $d$.  Eq. (S20) could be considered the main theoretical result of this work: it prescribes the dissipation dynamics of an anharmonic oscillator subject to dispersive loss. The equation governs the evolution of the entire density matrix of the anharmonic oscillator: not only the evolution of the populations (which are important for Fock state generation), but also the quantum coherences between different photonic states, which are important for monitoring the build-up and decay of phase and intensity correlations.  Eq. (S20) also serves as a foundation for the quantum Langevin description of nonlinear loss in systems with the Hamiltonian of Eq. (S2). This Langevin description enables us to study the quantum fluctuations of devices that use this nonlinear loss, such as lasers. For all of these reasons, the density matrix equation, Eq. (S20) provides the rigorous theoretical foundation for this work.

To make contact with the notations established in the main text (as well as more standard forms of the master equation), we will make the changes of definition $\kappa \rightarrow \kappa/2$ and $\gamma \rightarrow \gamma/2$. Additionally, we define the complex quantity $\mu_n = \frac{1}{2}\kappa - \frac{G_+G_-}{i(\omega_d - \omega_{n,n-1}) + \gamma/2}$. Eq. (S20) is then expressed as:
\begin{equation}
    \dot{\rho} = -\sum\limits_{n=0}^{\infty} n(\mu_n T_{n,n}\rho + \mu^*_n \rho T_{n,n} ) + \sum\limits_{m,n=0}^{\infty} \sqrt{m(n+1)}(\mu_m + \mu_{n+1}^*)T_{m-1,m}\rho T_{n+1,n},
\end{equation}
coinciding with Eq. (2) of the main text. 

\subsubsection{Equation of motion for photon probabilities}

The diagonal components of the density matrix $\rho_{n,n}$ correspond to the probability $p_n$ of there being $n$ photons in $a$. As the main text is primarily focused on realizing Fock and macroscopic sub-Poissonian states of light (with probability distributions more tightly concentrated than Poisson), the equation of motion for the photon probabilities plays a central role. Taking the $n,n$ matrix element of Eq. (S21), one immediately finds
\begin{equation}
    \dot{\rho}_{n,n} = -2n\text{Re }\mu_n \rho_{n,n} + 2(n+1)\text{Re }\mu_{n+1} \rho_{n+1,n+1},
\end{equation}
which is clearly of the form
\begin{equation}
    \dot{p}_{n} = -L_n p_n + L_{n+1}p_{n+1},
\end{equation}
with $L_n = 2n\text{Re }\mu_n$ found as:
\begin{equation}
    L_n =  n\left(\frac{\kappa\delta_n^2 + \gamma|\lambda|^2 + 2\sqrt{\kappa\gamma}\delta_n|\lambda|\cos\phi}{\delta^2_n + \gamma^2/4}\right)
\end{equation}
establishing Eqs. (3-4) of the main text (noting that $p(n) \equiv \rho_{n,n}$). 

The solution of Eq. (S23) provides the time-dependent probability distribution of $a$, giving access to all moments of the photon number operator. In many cases, we are primarily only interested in the dynamics mean and the variance. Thus, it is useful to derive an equation of motion for the mean and variance of the probability distribution. We shall do so in the approximation that the uncertainty $\Delta n$ is small compared to the mean $\bar{n}$, a statement which is almost always valid for states we consider, including Poissonian states (where $\Delta n = \sqrt{\bar{n}} \ll \bar{n}$ provided $\bar{n} \gg 1$). As a result of Eq. (S23), a general moment of the distribution $\langle n^k \rangle$ evolves according to
\begin{equation}
    \dot{\langle n^k \rangle} = -\sum\limits_{n=0}^{\infty} n^k L_n p_n + \sum\limits_{n=0} n^k L_{n+1}p_{n+1}.
\end{equation}
Shifting the index of the second term from $n+1 \rightarrow n$ and making use of the fact that $L_0 = 0$, we find 
\begin{equation}
    \dot{\langle n^k \rangle} = \langle \left((n-1)^k - n^k \right)L(n)\rangle,
\end{equation}
Thus, the mean evolves according to:
\begin{equation}
    \dot{\bar{n}} = -\langle L(n)\rangle,
\end{equation}
where we have denoted the mean as $\bar{n}$ to make contact with notations from the main text (other average quantities in this section will not get a bar). The second moment evolves according to: 
\begin{equation}
    \dot{\langle n^2 \rangle} = -\langle (2n-1)L(n)\rangle.
\end{equation}
The variance satisfies the equation of motion $\dot{(\Delta n)^2} = \dot{\langle n^2 \rangle} - 2\bar{n}\dot{\bar n}$. To proceed, we will consider distributions for which the distribution is sharply peaked about mean $\bar{n}$ (and is singly-peaked), such that  $\Delta n \ll \bar{n}$. In this case, we make a continuous approximation for the probability distribution: $p_n \rightarrow p(n)$, with averages given by $\langle f(n)\rangle = \int\limits_0^{\infty} dn ~f(n)p(n)$. Since the distribution is sharply peaked compared to the scale of variation of $L(n)$, we may Taylor expand the loss about the mean: $L(n) \approx L(\bar{n}) + (n-\bar{n})L'(\bar{n}) + \frac{1}{2}L''(\bar{n})(n-\bar{n})^2$. To lowest order, the mean simply evolves according to 
\begin{equation}
    \dot{\bar{n}} = -L(\bar{n}).
\end{equation}
Meanwhile, the variance is found as:
\begin{align}
  \dot{(\Delta n)^2} &= -\int\limits_0^{\infty} dn~ p(n) (2(n-\bar{n})-1)L(n) \nonumber \\ 
  &= -\int\limits_0^{\infty} dn~ p(n) (2(n-\bar{n})-1)\left(L(\bar{n}) + (n-\bar{n})L'(\bar{n}) + \frac{1}{2}L''(\bar{n})(n-\bar{n})^2 \right) \nonumber \\
  &= L(\bar{n}) - \left(2L'(\bar{n}) - \frac{1}{2}L''(\bar{n}) \right)(\Delta n)^2  + O((\Delta n)^3) \nonumber \\
  &\approx L(\bar{n}) - 2L'(\bar{n})(\Delta n)^2.
\end{align}
Here, we have used the simplification that $\langle n-\bar{n} \rangle = 0$. We have also ignored higher order variations in the distribution, and made a somewhat crude approximation that $4L' \gg L''$, which occurs when the distribution varies over a scale large compared to 1 (and hence is not perfectly accurate in the Fock-state regime). Still, the approximate equations capture the dynamics of the first two cumulants fairly well. The approximate equations for the cumulants, Eqs. (S29) and (S30) correspond to Eq. (6) of the main text.

\subsubsection{Equation of motion for field coherences}

Although we do not use this result in the main text, we expect that the equation of motion for the off-diagonal terms will play an important role in a theory of phase and higher-order coherence in the presence of nonlinear loss. Hence, we provide an explicit equation of motion for the $k$th coherence, corresponding to the off-diagonal components of the density matrix $\rho_{n-k,n}$. The equation of motion follows from Eq. (S21) as 
\begin{equation}
    \dot{\rho}_{n-k,n} = -((n-k)\mu_{n-k} + n\mu_n^*)\rho_{n-k,n} + \sqrt{(n-k+1)(n+1)}(\mu_{n-k+1}+\mu^*_{n+1})\rho_{n-k+1,n+1}.
\end{equation}

\subsection{Physical interpretation of the loss terms}

Let us now discuss the physical interpretation of the loss found in Eq. (S24). We shall take two approaches. In the first, we derive the Heisenberg equations of motion for this system, neglecting nonlinearity, and examine the mean-field limit.  We will show that the resulting model coincides with the so-called Friedrich-Wintgen model of two spatially co-located resonances with a common port. This model is known to support bound states in the continuum: modes that, although embedded in a reservoir of continuum states, have zero \cite{hsu2016bound}. This will be due to destructive interference (of the Fano type, between two different leakage pathways). In the second, we show that the loss is what would be expected from a mirror with a frequency dependent Fano reflectivity profile (by comparing to the standard classical model of Fano resonances).

\subsubsection{Connection to Fano interference and to bound states in the continuum}

We derive a Heisenberg equation of motion for $a$ and $d$ in the absence of nonlinearity. In the Appendix, we derive Heisenberg equations taking into account nonlinearity, and show that in the adiabatic approximation, identical conclusions are drawn (as compared to the density matrix treatment of the previous sections). From the Hamiltonian of Eq. (S2), the Heisenberg equations of motion for $a,d,b_k$ are given as:
\begin{align}
  \dot{a}  &= -i\omega_a a - i\lambda^* d - i\sum\limits_k g_k^* b_k \nonumber \\
  \dot{d}  &= -i\omega_d d - i\lambda a - i\sum\limits_k v_k^* b_k \nonumber \\
  \dot{b}_k &= -i\omega_k b_k - i(g_k a + v_k d).
\end{align}
To proceed, we will eliminate the reservoir. The formal solution to the reservoir equation of motion is given as
\begin{equation}
    b_k(t) = b_k(t_0)e^{-i\omega_k (t-t_0)} - i\int\limits_{t_0}^t dt' ~e^{-i\omega_k(t-t')}(g_k a(t') + v_k d(t')),
\end{equation}
with $t_0$ being the initial time (e.g., $t_0 = 0$ or $t_0 = -\infty$). Plugging this into the equation of motion for $a$ and $d$, and considering a white-noise reservoir $g_k = g$, $v_k = v$ (with both $g,v$ real), we have  
\begin{equation}
\begin{pmatrix}
\dot{a} \\
\dot{d}
\end{pmatrix} = \left[-i\omega_d -\begin{pmatrix}
i\delta + \frac{1}{2}\kappa && i\lambda^* + \frac{1}{2}\sqrt{\kappa\gamma} \\
i\lambda + \frac{1}{2}\sqrt{\kappa\gamma}  && \frac{1}{2}\gamma 
\end{pmatrix} \right]
\begin{pmatrix}
a \\
d
\end{pmatrix} + \begin{pmatrix}
F_a \\
F_d
\end{pmatrix}.
\end{equation}
Here, we have defined $\kappa = 2\pi \rho_0 g^2$ and $\gamma = 2\pi \rho_0 v^2$, with $\rho_0$ the density of continuum states. The terms $F_a$ and $F_d$ are operator-valued Langevin forces (Langevin forces will be elaborated on in the section ``Quantum Langevin theory of the nonlinear Fano resonance''). They have the property that for a vacuum reservoir, $\langle F_{a,d} \rangle = 0$. The non-zero second-order correlators, for a vacuum reservoir, are given as $\langle F_a(t)F^{\dagger}_a(t')\rangle = \kappa\delta(t-t')$,  $\langle F_d(t)F^{\dagger}_d(t')\rangle = \gamma\delta(t-t')$, and  $\langle F_a(t)F^{\dagger}_d(t')\rangle = \langle F_d(t)F^{\dagger}_a(t')\rangle = \sqrt{\kappa\gamma}\delta(t-t')$.

As discussed in the main text, much intuition can be built by examining the equation of motion for the mean values of $a,d$, which we denote as $A,D$. The equation of motion:
\begin{equation}
\begin{pmatrix}
\dot{A} \\
\dot{D}
\end{pmatrix} = \left[-i\omega_d -\begin{pmatrix}
i\delta + \frac{1}{2}\kappa && i\lambda^* + \frac{1}{2}\sqrt{\kappa\gamma} \\
i\lambda + \frac{1}{2}\sqrt{\kappa\gamma}  && \frac{1}{2}\gamma 
\end{pmatrix} \right]
\begin{pmatrix}
A \\
D
\end{pmatrix}.
\end{equation}
is simply Eq. (5) of the main text. Let us now diagonalize this matrix to isolate the coupled modes of the system. The two eigenvalues are found to differ considerably in overall scale (assuming $\kappa \ll \gamma$), one is $O(\gamma)$, while the other is $O(\kappa)$ (and the corresponding eigenvector is approximately $a$). The lower loss mode (which is $O(\kappa)$) has eigenvalue
\begin{equation}
    z = \frac{1}{4}\left(-\gamma-2i\delta-\kappa + \sqrt{(\gamma+2i\delta+\kappa)^2 - 4(2i\gamma\delta-4i\sqrt{\kappa\gamma}\text{Re }\lambda +4|\lambda|^2)} \right).
\end{equation}
In the limit of $\kappa, \lambda \ll \gamma$, we find that the real part of the eigenvalue is simply
\begin{equation}
    \text{Re }z = -\frac{1}{2} \frac{\kappa\delta^2 + \gamma|\lambda|^2 + 2\sqrt{\kappa\gamma}\delta\text{Re }\lambda}{\delta^2 + \gamma^2/4}.
\end{equation}
The associated temporal loss coefficient of the mode is simply $\kappa = -2\text{Re }z$, which coincides with the loss $L_n = n\kappa(n)$ Eq. (S24), except that the detuning is not $n$-dependent in Eq. (S37) (as we have not included nonlinearity). This comparison however makes it clear that the effect of nonlinearity is simply to control the value of $\delta$: stated operationally, the role of nonlinearity is to take $\delta \rightarrow \delta_n$.

Now, let us connect this result to the physics of Fano interference and the related phenomenon of bound states in the continuum. For certain values of the parameters ($\kappa,\gamma,\lambda$) in Eq. (S37), the loss can disappear. This is due to destructive interference of (1) a direct pathway for $a$ to leak out and (2) a pathway in which $a$ couples into $d$ before leaking out. To see more explicitly how the loss can vanish, consider the case of no direct coupling ($\lambda = 0$). Such an interference is known as Fano interference, as it can lead to an asymmetric lineshape in the presence of a non-zero $\lambda$. In this case, the numerator of Eq. (S37) is simply $\kappa\delta^2$, which vanishes for $\delta = 0$ (corresponding to the usual Fano transmission dip to be elaborated on in the next subsection). This mode, which has exactly zero loss, is known as a bound state in the continuum (BIC), which is of much recent interest in photonics (see e.g., \cite{hsu2016bound} for a review of the field). It is referred to as such because the cavity mode is localized (it does not leak), despite the existence of a reservoir of far-field modes for which this cavity mode can couple. 

These BICs can be shown to follow from exactly the classical model of Eq. (S35) (see Eq. 4 of \cite{hsu2016bound}), which is referred to as the Friedrich-Wintgen model \cite{friedrich1985interfering}, which is known to provide a simple model of BIC formation. Our quantum mechanical treatment of this system (in the linear case, as in Eq. (S34)) and in the nonlinear case is thus tantamount to a quantum theory of nonlinear bound states in the continuum, which appear to lead to Fock- and sub-Poissonian state generation. To our knowledge, such a quantum mechanical model, and these conclusions have not been previously reported.



\subsubsection{Interpretation of $d$-mode as a frequency-dependent mirror}

To get a further understanding of Eq. (S24), let us consider a related problem: the transmission and reflection of classical light scattering from a Fano mirror (a system with a Fano resonance). This problem has been studied by many authors, and is commonly considered in the field of nanophotonics. Consider a wave incident on a Fano mirror surrounded by air (e.g., a photonic crystal mirror). The wave has frequency $\omega$, the Fano mirror has frequency $\omega_0$, and radiative losses governed by the amplitude decay time $2/\gamma$ with $\gamma$ the energy decay rate. It can be shown \cite{fan2002analysis,fan2003temporal} that the energy transmission coefficient is then given by 
\begin{equation}
T = \frac{|t_d|^2\delta^2 + |r_d|^2\gamma^2/4 \pm |r_d t_d|\gamma\delta}{\delta^2 + \gamma^2/4},
\end{equation}
with $\delta = \omega-\omega_0$ and $r_d,t_d$ representing reflection and transmission coefficients associated with the \emph{direct} reflection and transmission of the incident light (i.e., without coupling into the internal mode of the mirror). These direct channels interfere with the indirect channel. Here, the $\pm$ denotes the case of an even/odd mode. Comparing this with Eq. (S24), we see that the losses are quite similar in form. In fact, we see that by taking Eq. (S24) and applying: $\omega \rightarrow \omega_{n,n-1}, \gamma \rightarrow \gamma, |t_d| \rightarrow \sqrt{\frac{2L\kappa}{c}}, |r_d| \rightarrow \sqrt{\frac{8L}{c\gamma}}|\lambda|$, with $L$ the length of the cavity supporting mode $a$, we have:
\begin{equation}
T_n \equiv T(\omega_{n,n-1}) = \frac{2L}{c}\frac{\kappa \delta_n^2 + \gamma |\lambda|^2  \pm 2\sqrt{\kappa\gamma}\delta_n|\lambda|}{\delta_n^2 + \gamma^2/4},
\end{equation}
which, stated differently, can be written as
\begin{equation}
L_n = n \times \left(\frac{cT_n}{2L} \right),
\end{equation}
for the case of $\phi = 0$ or $\pi$. This is Eq. (4) of the main text. Our model also considers more general coupling phases between the direct and indirect channels. 

Thus, the physical interpretation is evidently that the loss per photon ($L_n/n$) is simply the round-trip rate of light propagation in the cavity, multiplied by the cavity transmission. The mode $d$ acts as the resonance associated with a frequency-dependent end-mirror (this viewpoint is also described from a quantum mechanical density matrix model in \cite{vcernotik2019cavity}). This is largely what one intuitively expects, and is borne out from the density-matrix approach in the adiabatic approximation. This identification however, suggests a generalization to more complicated Fano mirrors, supporting perhaps multiple internal modes: the loss can be specified in terms of the experimental transmission as a function of frequency.

\subsection{Quantum Langevin theory of the nonlinear Fano resonance}

In this section, we develop a complementary perspective on the description of dissipation in a nonlinear resonator with sharply varying loss. In quantum optics, it is well-established that there are two often equivalent ways to describe dissipation. The first is by deriving a master equation for the density-matrix, as we have in the section titled ``Master equation of the nonlinear Fano resonance.'' The second is by deriving quantum Langevin (or Heisenberg-Langevin (HL)) equations for the Heisenberg-picture operators for the system. The quantum Langevin equations resemble classical equations that describe damping, except with operator-valued forces added to the equations to ensure preservation of operator commutation relations at all times. The two methods are complementary to each other, and each presents definite advantages over the other. In the density matrix approach, the equations for the density matrix elements are linear, and it is possible to find the evolution of the density matrix elements in a conceptually straightforward way. The density matrix method is the one which is mostly used in modern quantum engineering, and we have thus made the density-matrix approach the primary method. 

On the other hand, the Heisenberg-Langevin equations are generally nonlinear operator equations with quantum stochastic force terms that have no definite numerically implementable representation (though they may be mapped to classical stochastic differential equations which can then be solved). However, the main analytical advantages of the HL approach emerge in situations where quantum fluctuations are small compared to the mean values (as is the case in every system we analyze in the main text). In that case, operator expectation values, even for macroscopic states of light (that cannot be numerically stored as a density matrix, due to sheer dimensionality), can be readily found through a small number of coupled linear differential equations. From a fundamental standpoint, the Heisenberg-Langevin approach also has the advantage of bearing close similarity to classical equations of motion and thus providing a great deal of intuition. Very often, one may simply take classical equations, add stochastic force terms, and find the correlation functions of the forces through the so-called ``Einstein relation'' (as described in textbooks such as  \cite{yamamoto1999mesoscopic,chow2012semiconductor}). The Langevin approach has proven itself to be very useful in the context of laser physics for this reason. From the standpoint of lasers, it is also important because: for many important gain media, such as solid-state and semiconductor gain media, one cannot eliminate the gain from the density matrix, and thus cannot express the dynamics of the photon in terms of a time-local differential equation. Motivated by these advantages, we now develop the Heisenberg-Langevin equations for the photon number operator in a system with the nonlinear loss of Eq. (S24).

We follow the general method for deriving Langevin equations for quantum systems presented in Ref. \cite{haken1981waves} (there, the method is applied to derive Langevin equations for a two-level system). The method allows us to derive a Langevin equation in correspondence with the density matrix equation, Eq. (S21). Let us derive a Langevin equation to describe the evolution of the photon number operator, which is related to the photon probabilities, and thus the diagonal components of the density matrix.  In the method of Ref. \cite{haken1981waves}, one ``Langevinizes'' the density matrix equation, e.g., Eq. (S23), by assuming an equation of the form
\begin{equation}
    \dot{T}_{n,n} = -L_nT_{n,n} + L_{n+1}T_{n+1,n+1} + F_{n,n},
\end{equation}
where $T_{n,n} = |n\rangle\langle n|$ is a projector whose expectation value is simply $p_n$. The $F_{n,n}$ are operator valued Langevin forces associated with the quantum fluctuations which are concomitant with nonlinear dissipation. The force is stipulated to have zero mean but finite second-order correlations that are delta-correlated (schematically $\langle F(t)F(t')\rangle = 2D\delta(t-t')$ for some operator-valued ``diffusion coefficient'' $D$). Note that Eq. (S41) should be thought of as the operator equation in correspondence with Eq. (S23): taking the expectation value $\text{tr}[\rho \dot{T}_{n,n}] = \dot{p}_{n}$ yields $-L_n\rho_{n,n} + L_{n+1}\rho_{n+1,n+1}$ (using the fact that $\langle F_{n,n} \rangle = 0$). 

For a general Langevin equation of the form $\dot{A}_{\mu} = D_{\mu} + F_{\mu}$, where $A_{\mu}$ and $D_{\mu}$ are system operators, and $F_{\mu}$ is a Markovian Langevin force of zero mean $-$ quantum mechanical consistency (e.g., preservation of commutators) imposes a constraint on the correlation functions between different forces ($F_{\mu}, F_{\nu}$). In particular, the correlators must satisfy the so-called Einstein relation for the diffusion coefficient $D_{\mu\nu}$, defined such that $\langle F_{\mu}(t)F_{\nu}(t')\rangle \equiv 2\langle D_{\mu\nu}\rangle \delta(t-t')$. The Einstein relation reads \cite{chow2012semiconductor}:
\begin{equation}
    2\langle D_{\mu\nu}\rangle = \frac{d}{dt}\langle A_{\mu}A_{\nu}\rangle - \langle A_{\mu}D_{\nu}\rangle - \langle D_{\mu}A_{\nu}\rangle,
\end{equation}
As applied to Eq. (S41), the corresponding $D_{\mu}$ is $-L_nT_{n,n}+L_{n+1}T_{n+1,n+1}$ and the corresponding $F_{\mu}$ is $F_{n,n}$.

First, we find the diffusion coefficient $\langle D_{jj,kk}\rangle$, defined such that: \begin{equation}
\langle F_{j,j}(t)F_{k,k}(t')\rangle =  2\langle D_{jj,kk} \rangle \delta(t-t').
\end{equation}
It evaluates as:
\begin{align}
    2\langle D_{jj,kk} \rangle &= \delta_{j,k}\langle \dot{T}_{j,j}\rangle - \langle T_{j,j}D_{k,k}\rangle - \langle D_{j,j}T_{k,k} \rangle \nonumber \\
    &= \delta_{j,k}\langle -L_jT_{j,j} + L_{j+1}T_{j+1,j+1}\rangle + \langle (L_k \delta_{jk}T_{j,j} - L_{k+1}\delta_{j,k+1}T_{j,j})\rangle \nonumber \\
    &+ \langle (L_j\delta_{j,k}T_{j,j} - L_{j+1}\delta_{j+1,k}T_{j+1,j+1}) \rangle \nonumber \\
    &= \delta_{jk}(L_j\langle T_{j,j}\rangle + L_{j+1}\langle T_{j+1,j+1}\rangle) - \delta_{j,k+1}L_j\langle T_{j,j}\rangle - \delta_{j+1,k}\langle T_{j+1,j+1}\rangle.
\end{align}
As a sanity check on this result, consider the diffusion coefficient $2\langle D_{jj,jj}\rangle$. It evaluates as
\begin{equation}
    2\langle D_{jj,jj}\rangle = L_j\langle T_{j,j}\rangle + L_{j+1}\langle T_{j+1,j+1}\rangle = L_j p_j+ L_{j+1}p_{j+1}.
\end{equation}
In other words, the diffusion coefficient is the sum of the rate of transitions into and away from the state of $j$ photons. This property is a well-known result in the quantum theory of shot-noise \cite{lax1967quantum}, and indicates that the nonlinear loss dynamics can be thought of as associated with a type of \emph{nonlinear shot noise} whose added fluctuations depend on the number of photons present. 

The Einstein relation also enables us to specify a Langevin equation for the photon number operator itself (which is more readily measurable than the photon probabilities). The number operator is expressed in terms of projectors as $n = \sum\limits_{j=0}^{\infty} jT_{j,j}$. Therefore, we have 
\begin{align}
    \dot{n} &= \sum\limits_{j=0}^{\infty} -jL_jT_{j,j} + jL_{j+1}T_{j+1,j+1} + \sum\limits_{j=0}^{\infty} jF_{j,j} \nonumber \\ 
    &= \sum\limits_{j=0}^{\infty} -jL_jT_{j,j} + (j-1)L_{j}T_{j,j} + F_n \nonumber \\
    &= \sum\limits_{j=0}^{\infty} -L_jT_{j,j} + F_n \nonumber \\ 
    &= -\kappa(n)n + F_n,
\end{align}
where $L(n) = n\kappa(n)$ is understood to be a function of the $n$ operator. In this derivation, we have identified $F_n = \sum\limits_{j=0}^{\infty} jF_{j,j}$ and performed index manipulations similar to those used to derive Eq. (S26). Eq. (S46) is what one would write classically for a system with nonlinear loss, up to the Langevin force term $F_n$ \footnote{As a somewhat well-known example, such an equation would be used to describe the dynamics of the energy in a resonator with a saturable absorber (with or without Langevin forces \cite{haus1984waves,lathi1999influence})}.

The corresponding diffusion coefficient for $F_n$ may immediately be found from Eq. (S45). In particular:
\begin{align}
    2\langle D_{n,n}\rangle = &\sum_{j,k=0}^{\infty}2jk\langle D_{jj,kk}\rangle \nonumber \\ 
    =& \sum_{j,k=0}^{\infty} 2jk (\delta_{jk}(L_j\langle T_{j,j}\rangle + L_{j+1}\langle T_{j+1,j+1}\rangle) - \delta_{j,k+1}L_j\langle T_{j,j}\rangle - \delta_{j+1,k}\langle T_{j+1,j+1}\rangle) \nonumber \\
    =& \sum_{j=0}^{\infty} j(L_j \langle T_{j,j} \rangle - L_{j+1}\langle T_{j+1,j+1}\rangle) \nonumber \\
    =& \sum_{j=0}^{\infty} L_j\langle T_{j,j} \rangle = \langle n\kappa(n)\rangle.
\end{align}

Eqs. (S46) and (S47) represent the main result of the Langevin theory of decay of an anharmonic oscillator with intensity-dependent loss of the type resulting from nonlinear dispersive loss introduced in Sec. II. Although we have derived the Langevin equation from the density matrix equation specific  to the Hamiltonian of Eq. (S2) $-$ the content of Eqs. (S46-S47) is more general and are expected to describe photon number fluctuations in generic systems for which the loss coefficient depends on photon number.

Before moving on to the analysis of lasers employing this sharp loss, we comment on the ``Langevinization'' procedure. As stated, Eq. (S41) appears as an unjustified assumption (regardless of how well it works). We note that such an equation may also be more rigorously derived by considering an explicit reservoir, writing the Heisenberg equations of motion for $a$ and $d$, and integrating out the reservoirs in the Markov and adiabatic approximations. This is demonstrated in the Appendix, and in some cases provides a cleaner derivation of the nonlinear loss of Eq. (S24).

In the next section, we will use this Langevin equation, in conjunction with the standard Langevin equations describing a pumped gain medium, to derive the quantum statistical theory of lasers with sharp intensity-dependent loss. We then show how Fock and macroscopic sub-Poissonian states result. 

\section{Lasers based on sharply nonlinear loss}

In this section, we develop the quantum theory of lasers which employ the nonlinear loss leading to Fock- and sub-Poissonian state generation. We shall approach the problem in steps: first, discuss the system purely classically, in terms of rate equations for the population inversion and the cavity photon number. Then we convert these equations into Langevin equations, which will give information about fluctuations in the inversion and the cavity photon number. We derive the amplitude noise spectrum for the cavity, which tells us about frequency-resolved fluctuations in the photon number, as well as the overall photon number uncertainty. 

From there, we will discuss a particularly simple limit of the equations in which the inversion relaxation time is fast compared to the cavity decay. In that limit, the gain can be adiabatically eliminated, and a simple equation of motion may be derived for the cavity photon density matrix. Using this, we can derive a simple rule for the photon number fluctuations in terms of the value of the loss and its derivative at the steady-state, justifying Eq. (7) of the main text.

\subsection{Quantum Langevin theory of photon number fluctuations in a system with sharp loss}

As described in the beginning of the section, we start by reminding the reader of the classical analysis of the laser shown in Fig. 3b of the main text. We consider a single-mode cavity with nonlinear loss coefficient $\kappa(n)$ which interacts with a gain medium through emission and absorption. We consider a generic model of a gain medium: e.g., a four-level system in which the upper pumping level and the lower lasing level decay rapidly (through non-radiative processes). Thus, the only relevant populations in the equations are that of the lower pump level (the ground state) and the upper lasing level. Such conditions are well respected in many efficient gain media (as one example: solid-state gain media such as Nd:YAG). We also consider the limit in which the gain is approximately non-depleted, such that most of the population is in the lower pump level (the ground state). Under these conditions, it is very well known that the dynamical evolution (and steady-state) of the photon number and the gain are captured by the canonical rate equations (see any laser textbook, e.g., \cite{siegman1986lasers}). Denoting the inversion as $N$ and the photon number as $n$, we have:
\begin{align}
\dot{n} &= (R_{\text{sp}}N - \kappa(n))n \nonumber \\
\dot{N} &= \Lambda - \left(\gamma_{||} + R_{\text{sp}}n \right)N.   
\end{align}
Here, we have defined $R_{\text{sp}}$ as the rate of spontaneous emission of the gain medium into the cavity mode (which, up to a prefactor, satisfies $R_{\text{sp}} = f\sigma_{\text{st}}v/V$ with $f$ the filling fraction of the gain, $\sigma_{\text{st}}$ the stimulated cross section of the gain, $v = c/n_{\text{eff}}$ the speed of light in the cavity, and $V$ the cavity mode volume). We have also defined the pumping rate of the gain medium $\Lambda$ (sometimes expressed as $\gamma_{||}N_0$ with $\gamma_{||}$ the rate of population decay and $N_0$ the unsaturated inversion). We have neglected terms related to spontaneous emission in both equations in (48), as they will be negligible (even from a quantum noise perspective).

\subsubsection{Steady-state operating condition}

To start, it will be useful to find the steady-state operating point of the laser, obtained by setting the left-hand side of Eq. (S48) to zero. In that case, we have for the inversion:
\begin{equation}
    N = \frac{\Lambda}{\gamma_{||} + R_{\text{sp}}n},
\end{equation}
and for the photon number:
\begin{equation}
    \frac{R_{\text{sp}}\Lambda}{\gamma_{||} + R_{\text{sp}}n} = \frac{R_{\text{sp}}N_0}{1 + n/n_s} = \kappa(n),
\end{equation}
where we have defined the saturation photon number $n_s = \gamma_{||}/R_{\text{sp}}$.

For a generic loss function $\kappa(n)$, the equilibrium condition cannot be solved analytically. However, it is easy to understand graphically, by plotting the saturable gain and the loss and looking for the intersection points, as we have in Figs. 3d and 4b of the main text. From such graphical solutions, it is easy to appreciate that if $\kappa(n_0) = 0$ for some $n_0 \neq 0$, it implies the existence of a solution of the equations for the mean for any non-zero value of $\Lambda$. In particular, even if $R_{\text{sp}}N_0 < \kappa(0)$, which means the gain is less than the loss (and thus the system will not lase), a solution will still exist (typically for $n$ not very different from $n_0$). If the laser instead starts from a state with $n > n_0$ photons, it will move to this steady state. 

Eq. (S48) can be thought of the lowest-order description of the system in the mean-field approximation (e.g., replacing operators for the inversion and photon number by $c$-numbers). We now go beyond the mean-field approximation to find the fluctuations. 

\subsubsection{Quantum fluctuations}

Let us now find the quantum statistics of a nonlinear laser with sharp loss. The simplest treatment of quantum fluctuations in lasers proceeds by adding quantum Langevin forces to Eq. (S48) \cite{lax1967quantum}. In particular, we write:
\begin{align}
    \dot{n} &= \left(R_{\text{sp}}N - \kappa(n) \right)n + F_n \nonumber \\
    \dot{N} &= \Lambda - \left(\gamma_{||} +R_{\text{sp}}n \right)N + F_N.
\end{align}
The diffusion coefficients for the forces are given by:
\begin{align}
    2D_{nn} &= \langle \left(R_{\text{sp}}N + \kappa(n) \right)n \rangle \nonumber \\
    2D_{nN} &= 2D_{Nn} = -\langle R_{\text{sp}}Nn \rangle  \nonumber \\
    2D_{NN} &= \Lambda + \langle \left(\gamma_{||} +R_{\text{sp}}n \right)N \rangle.
\end{align}
Compared to previous Langevin treatments of fluctuations in lasers (see as examples \cite{lax1967quantum, yamamoto1992photon, yamamoto1999mesoscopic, van2000laser}), the only difference is the presence of the nonlinear loss $\kappa(n)$. The remarkable statement is that when $\kappa(n)$ takes the form implied by Eq. (S24), extremely sub-Poissonian states, approaching Fock states can result (with far lower noise than allowable by the types of low-order nonlinearities studied previously  \cite{walls1990amplitude,ritsch1990quantum,wiseman1991noise,ritsch1992quantum}). 

We now solve for the photon statistics. We are primarily interested in the cavity photon statistics at the steady-state operating point of the laser (thus we will not consider their evolution in time starting from vacuum). We will quantify the photon statistics primarily by the mean and variance of the cavity photon number (with a variance of zero corresponding to a cavity Fock state). In all cases we consider in this paper (even the noisiest ones), the quantum fluctuations of the photon number and inversion are small compared to the mean values. Thus, we may linearize the Langevin equations (which are nonlinear in $n$ and $N$) around their mean values as: $n = \bar{n} + \delta n$ and $N = \bar{N} + \delta N$. The quantities $\bar{n}$ and $\bar{N}$ are c-number (mean) values (given by Eqs. (S49) and (S50)) while $\delta n$ and $\delta N$ are operator-valued fluctuations. It follows immediately from the definitions above, and the zero mean-values of the forces, that $\langle n\rangle = \bar{n}$ and $(\Delta n)^2 = \langle (\delta n)^2\rangle$. These fluctuations are of the same order as the Langevin forces $F_n$ and $F_N$. 

The fluctuations of the photon number and inversion satisfy the pair of coupled equations
\begin{equation}
\begin{pmatrix}
\dot{\delta n} \\
\dot{\delta N}
\end{pmatrix} = \begin{pmatrix}
-\kappa'(\bar{n})\bar{n} && R_{\text{sp}}\bar{n} \\
 -R_{\text{sp}}\bar{N} && -\left( \gamma_{||} + R_{\text{sp}}\bar{n} \right)
\end{pmatrix}
\begin{pmatrix}
\delta n \\
\delta N
\end{pmatrix} + 
\begin{pmatrix}
F_n \\
F_N
\end{pmatrix}.
\end{equation}
Here, we have introduced $\kappa'(n) = d\kappa/dn$, which quantifies the sharpness of the loss. To solve this equation, it is convenient to Fourier transform (defining e.g., $\delta n(t) = \int\limits_{-\infty}^{\infty}\frac{d\omega}{2\pi}~e^{-i\omega t}\delta n(\omega)$). The uncertainty in the photon number then follows as $(\Delta n)^2 = \int\limits_{-\infty}^{\infty}\frac{d\omega}{2\pi}~S_{nn}(\omega)$, with $S_{nn}(\omega) = \langle \delta n^{\dagger}(\omega) \delta n(\omega)\rangle$ being the cavity photon amplitude noise spectrum. The Fourier transformed equations read:
\begin{equation}
\begin{pmatrix}
i\omega -\kappa'(\bar{n})\bar{n} && R_{\text{sp}}\bar{n} \\
 -R_{\text{sp}}\bar{N} && i\omega -\left( \gamma_{||} + R_{\text{sp}}\bar{n} \right)
\end{pmatrix}
\begin{pmatrix}
\delta n (\omega) \\
\delta N (\omega)
\end{pmatrix} = 
-\begin{pmatrix}
F_n (\omega) \\
F_N (\omega)
\end{pmatrix}.
\end{equation}
This admits the solution:
\begin{equation}
\begin{pmatrix}
\delta n (\omega) \\
\delta N (\omega)
\end{pmatrix} = -\frac{1}{(\Omega^2-\omega^2)-i\omega\eta}\begin{pmatrix}
i\omega - \Gamma && -R_{\text{sp}}\bar{n} \\
R_{\text{sp}}\bar{N} && i\omega -\kappa'(\bar{n})\bar{n} 
\end{pmatrix}
\begin{pmatrix}
F_n (\omega) \\
F_N (\omega)
\end{pmatrix},
\end{equation}
where we have defined $\Gamma = \gamma_{||}+R_{\text{sp}}\bar{n}$, as well as  the ``relaxation oscillation frequency''
\begin{equation}
\Omega^2 = \left(\Gamma\kappa'(\bar{n}) + R_{\text{sp}}\kappa(\bar{n}) \right)\bar{n},
\end{equation}
and the ``relaxation oscillation damping rate''
\begin{equation}
\eta = \Gamma + \kappa'(\bar{n})\bar{n}.
\end{equation}
With these definitions, the photon number spectrum evaluates as:
\begin{equation}
    S_{nn}(\omega) = 2\kappa(\bar{n})\bar{n}\times \frac{\omega^2 + \Gamma^2}{(\omega^2 - \Omega^2)^2 + \omega^2\eta^2}.
\end{equation}
Noise spectra of this form are plotted in Fig. 4 of the main text.

\subsection{Quantum statistics of lasers with sharp loss for fast inversion lasers}

While the Langevin framework gives access to the fluctuations of the photon number in the steady-state, it is much less simple to acquire dynamical information regarding the probability distribution of the photon number (as well as higher-order moments of the distribution). It becomes possible to find explicitly a simple, temporally local equation of motion for the density matrix of the cavity photon as a function of time in the limit where the population decay of the gain medium $\gamma_{||}$ is fast compared to the cavity lifetime. 

This so-called ``class A regime'' of laser operation often holds in gain media such as gases and molecular dyes. However, the inequality depends on the cavity lifetime, which can be made large using a long cavity or highly reflective mirrors. Thus in principle, semiconductor gain media can also behave as ``class A'' systems (for example, in external cavity configurations) -- and even rare earth gain media in principle could (using cavities formed via crystalline supermirrors). 

In this limit, the gain medium can be fully adiabatically eliminated. The resulting laser theory is called the Lamb-Scully theory of the laser \cite{scully1967quantum, scully1999quantum}. Let us now write down an equation of motion for the cavity photon density matrix for a system with gain and nonlinear loss. The contribution of the gain medium to the density matrix equation of motion is well-known from the Lamb-Scully theory, and so we merely quote the answer below. The loss terms of Eq. (S23) can simply be added to the contributions from the gain, as the photon state (which changes on the cavity time-scale) hardly changes over the time-scale $\gamma_{||},\gamma_{\perp}$. 

The combined effect of the gain-medium and the cavity loss on the equation of motion for the photon probabilities is
\begin{equation}
    p_{n} = A_n p_{n-1} - (A_{n+1}+L_n)p_n + L_{n+1}p_{n+1},
\end{equation}
where
\begin{equation}
    A_n = \frac{An}{1+n/n_s},
\end{equation}
with $A$ the linear gain coefficient and $n_s$ the saturation photon number. Here, we have assumed that the gain medium is resonant with the cavity. Few qualitative changes are introduced by including a finite detuning. Note that the value of $A$ which ensures consistency with Eq. (S49) is $A = R_{\text{sp}}N_0$. 

The steady-state photon statistics are found by setting $\dot{p}_{n} = 0$ with the normalization constraint $\sum\limits_n p_{n} = 1$. In steady-state, $\dot{p}_{n} = 0$ implies 
\begin{equation}
    A_n p_{n-1}  - L_{n}p_{n}  = A_{n+1}p_{n} - L_{n+1} p_{n+1}.
\end{equation}
Defining the difference $S_n = A_n p_{n-1}  - L_{n}p_{n}$, we see that $S_n = S_{n+1}$. Since $S_0 = A_0p_{-1} - L_{0}p_0 = 0$, we have that $S_n = 0$ for all $n$, and thus the simpler recursion relation:
\begin{equation}
  p_{n+1} = \frac{A_{n+1}}{L_{n+1}}p_{n} \implies p_n = \frac{1}{Z}\left(\prod\limits_{m=1}^n \frac{A_m}{L_{m}}\right) \equiv \frac{1}{Z}\left(\prod\limits_{m=1}^n \frac{G_m}{\kappa_{m}}\right),
\end{equation}
with $Z$ a normalization constant enforcing $\sum\limits_n p_{n} = 1$. We have also expressed the distribution in terms of the temporal gain coefficient $G_n$ and temporal loss coefficient $\kappa_n$. Using this form for the probability distribution, we find an analytical approximation for the photon number uncertainty. We consider distributions which are singly-peaked and vary on a scale large compared to one (making the approximation crude in the Fock state limit, but the resulting approximation is qualitatively predictive, even in that regime). Under these assumptions, we may make a continuum approximation for the probability distribution as follows. Express the probability distribution as
\begin{equation}
    p_n = \frac{1}{Z}\exp\left[\sum\limits_{m=1}^n \ln r_m \right],
\end{equation}
where $r_m = G_m/\kappa_m$. The peak of the distribution occurs for $\bar{n}$ such that $G_{\bar{n}} = \kappa_{\bar{n}}$. Physically, this is clear because it is the point at which gain balances loss. Mathematically, this is clear because for $m<\bar{n}$, $G > \kappa$ ($r > 1$) and the distribution is increasing (see Fig. 3 of main text for graphical ``proof'' of this statement). While for $m > \bar{n}$, $G < \kappa$ ($r < 1$) and the distribution is decreasing. Linearizing $r$ about the equilibrium point as $r(n) = 1 + r'(\bar{n})(n-\bar{n})$, such that $\ln r(n) \approx r'(\bar{n})(n-\bar{n})$, and making the continuum approximation for the distribution, Eq. (S63) may be approximated as
\begin{equation}
    p_n \approx \frac{1}{Z}\exp\left[\int\limits_{\bar{n}}^n dm~ r'(\bar{n})(m-\bar{n})\right] = \frac{1}{Z}\exp\left[-\frac{1}{2}|r'(\bar{n})|(n-\bar{n})^2\right],
\end{equation}
where in the last equality, we have used that $r' < 0$ (otherwise the equilibrium is not stable). From this expression, it immediately follows that the variance in the photon number is given as
\begin{equation}
    (\Delta n)^2 = \frac{1}{-\frac{d}{dn}\frac{G(n)}{\kappa(n)}\Big|_{\bar{n}}}.
\end{equation}
This establishes Eq. (7) of the main text. Note that for cases where the loss is sharp compared to the gain, we may ignore the derivative of $G$ and evaluate:
\begin{equation}
    (\Delta n)^2 \approx \frac{1}{\frac{G(\bar{n})\kappa'(\bar{n})}{\kappa^2(\bar{n})}} = \frac{1}{\kappa'(\bar{n})/\kappa(\bar{n})}.
\end{equation}
This equation shows that the fluctuations in the photon number are reduced when the loss is sharp compared to its equilibrium value (the latter of which is small near the zero of the loss of Eq. (S24)). 

\section{Numerical evidence for the effects predicted in the manuscript} 

In this section, we provide numerical results based on exact numerical time-dependent solutions, as well as exact numerical steady-states of the Liouvillian, to support the analytical results developed in this SI. The purpose of these numerical demonstrations is to provide a type of ``proof'' that the simple nonlinear Hamiltonian of Eq. (S2) indeed supports Fock-state (or approximate Fock-state generation), and to prove that laser action can indeed generate such unusually low noise states $-$ without resorting to the approximations (e.g., adiabatic elimination) that lead to our analytical theory (which more or less reaches the same conclusions).

\subsection{Numerical validation of transient noise condensation} 

To numerically demonstrate that the nonlinear coupled cavity Hamiltonian (Eq. (S2)) supports transient noise condensation similar to our analytical theory, we will numerically solve the master equation for the nonlinear Fano resonance (Eq. (S8)). Compared to the analytical theory, we do not assume the adiabatic approximation in the numerical solutions. Because we are numerically time-evolving an open system according to a Liouvillian, $-$ which has $N^4$ elements in its matrix representation ($N$ being the Hilbert space dimension) $-$ it is time-consuming to do simulations for large Fock states. Thus we demonstrate a "toy" example in which a 30-photon optical Fock state results (already such simulations take nearly two hours). The evaluation of the Liouvillian and the solution of the time-dependent equation of motion are performed in a standard numerical quantum optics package: in this case, QuantumOptics in the Julia programming language. Example code is provided \footnote{A github repository containing codes used to numerically validate the transient noise condensation and Fock lasing effects is here: https://github.com/nrivera494/photon-noise-condensation.}.

The results are shown in Fig. S1: there is a strong resemblance between Fig. S1(b) and Fig. 2 of the main text. An initially Poisson distribution condenses its noise by orders of magnitude, approaching a near 30-photon Fock state (corresponding with the zero of the loss of Eq. (S24)) with near unity probability. At the final time of the simulation, the probability of ending up with an optical 30-photon Fock state is 96\%. 

There is a somewhat apparent discrepancy when comparing the cumulants (Fig. S1(c)) to Fig. 2 of the main text.  The probability distribution at the final time is more sharply peaked around $n=30$ than at earlier times. But the Fano factor is higher (and in fact, appears to be quite high (about 0.5), indicating a somewhat modest noise reduction). This happens because it appears that a small part of the probability distribution, for lack of a better word, "tunnels" through the zero of the loss. One can see that the probability of being in the vacuum state increases over time (to a small value). In other words, the system displays some signature of bistability: the vast majority of the state is in the 30-photon Fock state while a very small part is in the vacuum state. This bimodality makes the uncertainty a poor indicator of the behavior of the distribution: it is sufficiently clear that the probability of generating a large Fock state in this system is quite high.

This bistability is somewhat unsurprising since the loss has zeros in two places (0 and $n_0$), indicating two valid steady states. We speculate that small (e.g., second-order) corrections away from adiabatic elimination could cause this (but we do not yet conclusively know what terms cause this). Nevertheless, the state already demonstrated through these simulations were mostly intended for ``proof purposes,'' would represent both the highest optical Fock state realized (by over an order of magnitude), and with a very high fidelity. It is likely that changes in parameters can improve this (since we made no attempt to optimize this).

\begin{figure*}[h]
    \centering
    \includegraphics[width=0.85\textwidth]{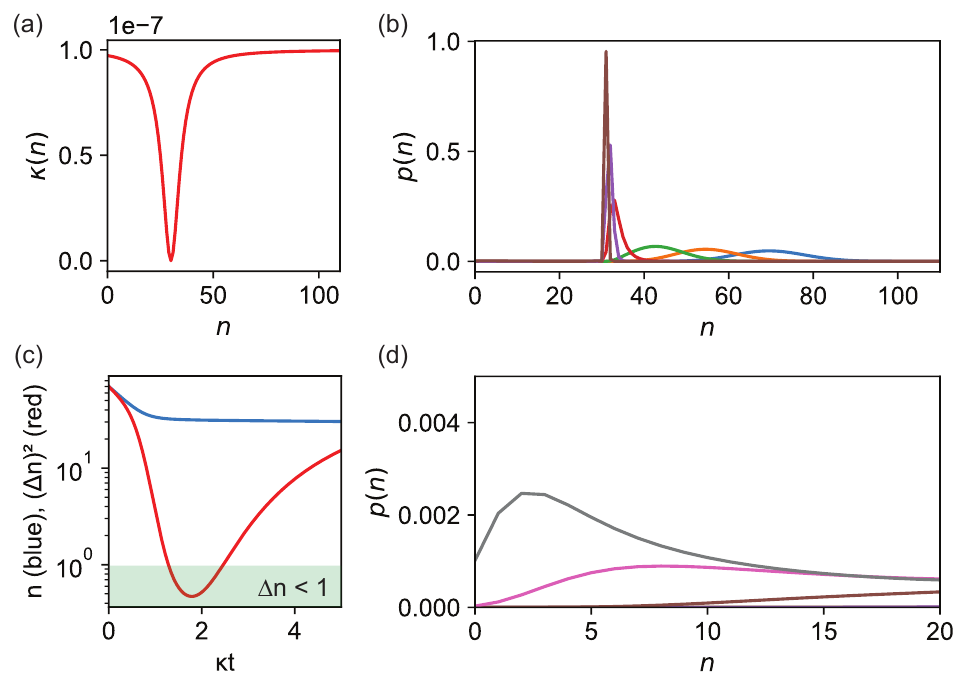}
    \caption{\textbf{Numerical demonstration of transient noise condensation from Eq. (S8).} (a) Temporal loss coefficient as a function of photon number. (b) Evolution of the photon statistics in $a$ for different times, assuming an initial Poisson distribution with 70 photons (blue curve). The dynamics largely mirror those presented by the analytical theory in Fig. 2 of the main text. (c) Mean and variance as a function of time, indicating the region where the photon distribution has an uncertainty less than 1. (d) Zoom-in of the small-photon number part of the distribution, showing that part of the distribution moves to smaller photon numbers, somewhat ``artificially'' diluting the Fano factor. The overall fidelity of generating a 30-photon Fock state in this example is 96\%. The parameters for the nonlinear system are $\beta = 5 \times 10^{-4}$, $\kappa = 10^{-7}$, $\gamma = 10^{-2}$, $\lambda = 0$, and $\omega_d = (1+\delta)\omega_a$ with $\delta = -3\gamma$.}
\end{figure*}

\subsection{Numerical validation of Fock lasing}

In this subsection, we demonstrate numerically (from steady-state solutions of the density matrix equation of motion) that a gain medium, coupled to the nonlinear coupled-resonator system, lases into a heavily sub-Poissonian state, approaching a Fock state. While it is essentially impossible to model from quantum mechanical first principles the interaction of $N$ pumped atoms with a cavity (because the Hilbert space dimension of $N \gg 1$ atoms is simply too large), it is possible to consider a related problem whose solution is representative of a many-body gain medium. In particular, we look at the coupling of a single pumped emitter
interacting with the cavity, and look at the photon probability distribution in the steady state. As in the previous subsection, the numerical calculations are performed in Julia's QuantumOptics package. Example code is provided.

Such a system, a single emitter coupled to a cavity (with a suitably rescaled coupling), is capable of correctly modeling the quantum dynamics of a laser, because in a laser, there are negligible inter-atom correlations (although there are implicit correlations in so far as all the atoms couple to the common cavity field that they interact with). As a result, as shown in Ref. \cite{scully1999quantum}, the resulting density matrix equations of motion for the system of cavity and gain medium are identical to that arising from the coupling of a single gain atom to the cavity (although of course, a single gain atom provides a much smaller amount of gain). 

The type of model considered here, of a single gain atom coupled to a cavity, beyond being useful for modeling purposes, also has a physical "life of its own." In particular, experiments exist demonstrating "one-atom lasing / masing" in which a single pumped atom is sufficient to exceed the threshold of the system (due to the very low losses of the system) \cite{mckeever2003experimental,an1994microlaser,liu2015semiconductor}. Such one atom lasers have been developed at both optical (with atoms coupled to high-finesse cavities) and microwave frequencies (with superconducting qubits).

The Hamiltonian of a four-level atom (states $1,2,3,4$ with a lasing transition $2-3$) coupled to the nonlinear cavity is given by
\begin{equation}
    H_{\text{laser}} = H_{ad} + \sum\limits_{\alpha=1}^4 E_{\alpha}|\alpha\rangle\langle \alpha| + \hbar g (\sigma^{+}a + a^{\dagger}\sigma^{-}),
\end{equation}
with $H_{ad}$ the Hamiltonian of Eq. (S3), $\sigma^+ = \sigma_{32} \equiv |3\rangle\langle 2|$, and $\sigma^- = \sigma_{23} = |2\rangle\langle 3|$. The atomic states are labeled in increasing energy order ($1$ is the ground state, $2$ is the lower lasing level, $3$ is the upper lasing level, and $4$ is the upper pump level). Here, we have not written the reservoir terms corresponding to cavity damping, atomic damping, and atomic pumping. We will consider them as contributing Lindblad terms to the equation of motion for the density matrix.  

The Lindblad term for the cavity, according to Eq. (S8) is $\mathcal{D}[X]$, with $X = \sqrt{\kappa}a + \sqrt{\gamma}d$ $\mathcal{D}[J] \equiv -\frac{1}{2}(J^{\dagger}J\rho + \rho J^{\dagger}J - 2J\rho J^{\dagger})$ being the standard dissipator for jump operator $J$. Defining $\sigma_{ij} = |i\rangle\langle j|$, the atomic damping terms are as follows: 
\begin{enumerate}
    \item The atom is pumped from $1$ to $4$ at rate $\Lambda$, with jump operator $\sigma_{41}$.
    \item The upper pump level $4$ decays to the lower lasing level $3$ at rate $\gamma_{34}$, with jump operator $\sigma_{34}$.
    \item The upper lasing level decays to the lower lasing level with relaxation time $\gamma_{||}$, with jump operator $\sigma_{23} = \sigma^{-}$.
    \item The lasing transition is subject to dephasing at rate $\gamma_{\perp}$ with jump operator $\sigma_z = \sigma_{33} - \sigma_{22}$.
    \item The lower lasing level decays to the ground level at rate $\gamma_{12}$ with jump operator $\sigma_{12}$.
\end{enumerate}

The Liouvillian operator $\mathcal{L}$ such that $\dot{\rho} = \mathcal{L}\rho$ is then given as
\begin{equation}
    \mathcal{L}\rho = -\frac{i}{\hbar}[H_{\text{laser}},\rho] + (\mathcal{D}[X] + \Lambda \mathcal{D}[\sigma_{41}] + \gamma_{34}\mathcal{D}[\sigma_{34}] + \gamma_{\perp}\mathcal{D}[\sigma^-] + \gamma_{||}\mathcal{D}[\sigma_{z}] + \gamma_{12}\mathcal{D}[\sigma_{12}])\rho.
\end{equation}

The steady state density matrix $\rho_{ss}$ is then found as the null eigenvector of the Liouvillian $\mathcal{L}\rho_{ss} = 0$. Thus, for a given set of parameters describing the laser system, we numerically implement the Liouvillian and find its zero eigenvalue. The steady-state density matrix is then used to calculated the photon probability distribution of $a$ from which the mean photon number, variance, and Fano factor are calculated. This is done as a function of the pump strength, and the results are presented in Fig. S2. 

In Fig. S2a, we plot a gain/loss curve similar to the ones employed in the main text (Figs. 3, 4). This will line-up well with the different regimes of operation (sub-threshold, bistable, near-Fock). Note that the agreement with the analytical theory of the SI is imperfect because at these low photon numbers, spontaneous emission affects the threshold. Nevertheless, the effects shown in the manuscript are all clearly present below (especially the very low-noise steady-states). Namely, we see that: after a threshold, the photon number starts to become significant and the system passes through a series of high-noise states into a low-noise state. The lowest relative noise on the input-output curve of panels (b,c) is just after the threshold, as is the case in Fig. 3 of the main text. In this case, it corresponds to a noise 90\% below the coherent-state limit (and much below what standard nonlinear absorbers and low-order nonlinearities provide). The photon uncertainty is about 1.9. The photon number is 35, which is near the approximate zero of the loss at 30 shown in panel (a) (it is higher for reasons that are evident from the gain-loss curves). 

Although we do not plot them here, we point out that the atomic populations are what one would expect from a canonical four-level gain: the lowest level is negligibly depleted, the lower lasing and upper pump levels have negligible population, and the small population in the upper lasing level is enough for inversion. Moreover, as in the previous subsection, the $d$ mode has very few photons in it, as expected from $\gamma \gg \kappa$ $-$ validating the assumptions underlying the adiabatic elimination of $d$ in the analytical theory.

\begin{figure*}[h]
    \centering
    \includegraphics[width=1\textwidth]{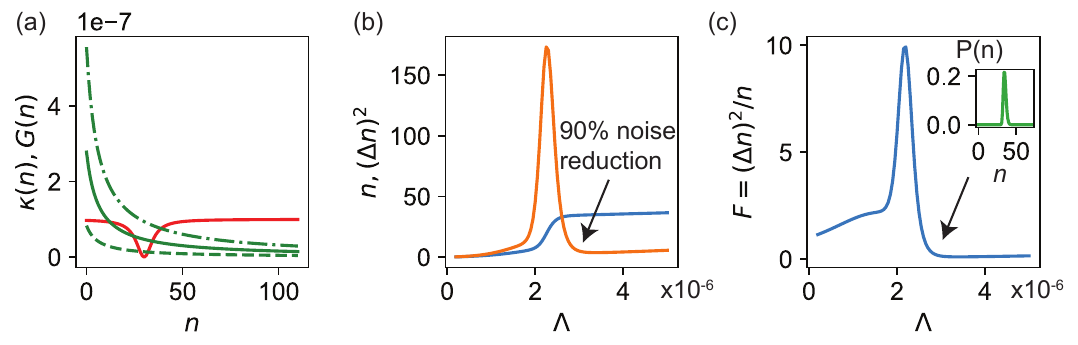}
    \caption{\textbf{Numerical demonstration of Fock lasing from steady state of the Liouvillian.} (a) Loss (red) and gain (green) curves for different values of the pump strength. (b) Mean number of photons in the cavity (blue), as well as variance (orange), as a function of pumping rate from the ground state to the upper pumping level. After a threshold, the photon number increases linearly, before going through a nearly discontinuous jump to a low noise state, with noise here 90\% below the coherent state level.  (c) Fano factor corresponding to the mean and variance in (b), with inset showing the photon probability distribution at the lowest-relative-noise point. Parameters for the nonlinear cavity are the same as Fig. S1 here. Parameters for the gain are: $g = 3 \times 10^{-4}$, $\gamma_{\perp} = 10^{-2}, \gamma_{||} = 10^{-4}, \gamma_{12} = 10^{-3}, \gamma_{34} = 1$ (exact value of $\gamma_{34}$ has little influence insofar as it is much faster than $\Lambda$ (all units are in units of the frequency of $a$ (e.g., 1.5 eV).}
\end{figure*}

\section{Summary of main results}

For ease of quotation, we compile in this section the main new equations derived in this work. 
\textbf{Master equation for a nonlinear resonance and a linear resonance coupled to a common continuum.} In the adiabatic approximation, where the damping rate of the nonlinear resonance is much smaller than that of the linear resonance, the equation of motion $\rho$ for the density matrix of the nonlinear resonance is given by:
\begin{equation}
    \dot{\rho} = -\sum\limits_{n=0}^{\infty} n(\mu_n T_{n,n}\rho + \mu^*_n \rho T_{n,n} ) + \sum\limits_{m,n=0}^{\infty} \sqrt{m(n+1)}(\mu_m + \mu_{n+1}^*)T_{m-1,m}\rho T_{n+1,n},
\end{equation}
with $\mu_n = \frac{1}{2}\kappa - \frac{G_+G_-}{i(\omega_d - \omega_{n,n-1}) + \gamma/2}$ and $T_{m,n} = |m\rangle\langle n|$. In this equation $n$ indexes over photon number in the nonlinear resonance with annihilation operator $a$ with frequency $\omega_a$, decay constant $\kappa$, and Kerr nonlinearity of strength $\beta$. The linear mode with annihilation operator $d$ has frequency $\omega_d$ and decay constant $\gamma \gg \kappa$. The term $G_+ = i\lambda^* + \frac{1}{2}\sqrt{\kappa\gamma}$ while $G_- = i\lambda + \frac{1}{2}\sqrt{\kappa\gamma}$. The frequency $\omega_{n,n-1} = \omega_a(1+2\beta n)$ is the intensity-dependent resonance frequency of the cavity.
\\
\textbf{Equation of motion for the probabilities.} The probability $p_n$ of $n$ photons being in the nonlinear resonance, $a$, evolve according to:
\begin{equation}
    \dot{p}_{n} = -L_n p_n + L_{n+1}p_{n+1},
\end{equation}
with $L_n$ found as:
\begin{equation}
    L_n =  n\left(\frac{\kappa\delta_n^2 + \gamma|\lambda|^2 + 2\sqrt{\kappa\gamma}\delta_n|\lambda|\cos\phi}{\delta^2_n + \gamma^2/4}\right),
\end{equation}
with $\kappa, \gamma, \lambda$ being defined above. The term $\delta_n = \omega_{n,n-1} - \omega_d$. We also define the temporal loss coefficient as $\kappa(n) = L_n/n$.
\\
\textbf{Equation of motion for $k$-th coherences of the field.} The off-diagonal components of the density matrix $\dot{\rho}_{n-k,n}$, with $k$ an integer, evolve according to:
\begin{equation}
    \dot{\rho}_{n-k,n} = -((n-k)\mu_{n-k} + n\mu_n^*)\rho_{n-k,n} + \sqrt{(n-k+1)(n+1)}(\mu_{n-k+1}+\mu^*_{n+1})\rho_{n-k+1,n+1}.
\end{equation}
\\
\textbf{Langevin equation for a nonlinear resonance.} The photon number operator $n$ in a nonlinear cavity with the loss of Eq. (S24) evolves according to the Langevin equation
\begin{equation}
    \dot{n} = -\kappa(n)n + F_n,
\end{equation}
where $\kappa(n)$ is the temporal loss coefficient defined earlier in this section, and $F_n(t)$ is a quantum Langevin force. The Langevin force has zero mean ($\langle F_n\rangle = 0$), and the diffusion coefficient of $F_n$ (defined so that $\langle F_n(t)F_n(t')  \rangle = 2\langle D_{n,n}\rangle \delta(t-t')$) is given by
\begin{equation}
2\langle D_{n,n}\rangle =  \langle n\kappa(n)\rangle.
\end{equation}
\\
\textbf{Noise spectrum of a Fock laser.} The spectrum of fluctuations for the cavity photon number, $S_{nn}(\omega)$ is defined such that the photon number variance $(\Delta n)^2 = \int\limits_{-\infty}^{\infty} \frac{d\omega}{2\pi}~ S_{nn}(\omega)$. The spectrum of fluctuations for a Fock laser $-$ for a four-level gain medium with fast decays of the upper pump and lower lasing level $-$ is given by:
\begin{equation}
    S_{nn}(\omega) = 2\kappa(\bar{n})\bar{n}\times \frac{\omega^2 + \Gamma^2}{(\omega^2 - \Omega^2)^2 + \omega^2\eta^2}.
\end{equation}
Here, $\bar{n}$ is the mean photon number in the laser cavity at steady-state, and $\Gamma = \gamma_{||} + R_{\text{sp}}\bar{n}$ with $\gamma_{||}$ the relaxation rate of the upper pump level and $R_{\text{sp}}$ the rate of spontaneous emission into the cavity mode. We have also for simplicity defined the ``relaxation oscillation frequency''
\begin{equation}
\Omega^2 = \left(\Gamma\kappa'(\bar{n}) + R_{\text{sp}}\kappa(\bar{n}) \right)\bar{n},
\end{equation}
and the ``relaxation oscillation damping rate''
\begin{equation}
\eta = \Gamma + \kappa'(\bar{n})\bar{n}.
\end{equation}
The term $\kappa'(\bar{n})$ is defined as $\frac{d\kappa}{dn}\Big|_{\bar{n}}$.
\\
\textbf{Photon probability distribution of a Fock laser.} The probability of $n$ photons being in the laser cavity, in the class A limit (where $\gamma_{\perp}, \gamma_{||} \gg \kappa$) is given as:
\begin{equation}
  p_{n} = \frac{1}{Z}\left(\prod\limits_{m=1}^n \frac{R_{\text{sp}}N_0}{(1+m/n_s)\kappa(m)}\right),
\end{equation}
with $Z$ a normalization constant and $N_0 = \Lambda/\gamma_{||}$ the unsaturated inversion, with $\Lambda$ the pumping rate of the upper lasing level. We have also defined the saturation photon number $n_s = \gamma_{||}/R_{\text{sp}}$.
\\
\textbf{Effect of gain and loss sharpness on photon uncertainty in the Fock laser.} The uncertainty of the photon number in the cavity, in the class A regime, is given by:
\begin{equation}
    (\Delta n)^2 = \frac{1}{-\frac{d}{dn}\frac{G(n)}{\kappa(n)}\Big|_{\bar{n}}},
\end{equation}
with $G(n)$ the intensity-dependent temporal gain coefficient. For cases where the loss varies much more sharply compared to the gain, $(\Delta n)^2$ may be approximated as:
\begin{equation}
    (\Delta n)^2 \approx \frac{1}{\frac{G(\bar{n})\kappa'(\bar{n})}{\kappa^2(\bar{n})}} = \frac{1}{\kappa'(\bar{n})/\kappa(\bar{n})}.
\end{equation}

\section{Potential extensions of the theory}

Here, we list some potential theoretical areas of exploration that should be enabled by the results here (experimental directions are discussed in the main text).

\begin{enumerate}
    \item The equation of motion for the density matrix, Eq. (S21), provides a starting point for many investigations of systems with nonlinear frequency-dependent loss. For example, one may use this equation to study statistics under coherent driving.
    \item The master equation of Eq. (S21), applied to describe coherence, also enables the study of the dynamical evolution of field- ($g^{(1)}(t,t')$), intensity- ($g^{(2)}(t,t')$) and higher-order field correlations ($g^{(k)}(t,t')$). The $k$-th order correlation functions are connected to the equation of motion for $\rho_{n-k,n}$ by the quantum regression theorem \cite{scully1999quantum}.
    \item The system introduced in this work, with Hamiltonian given by Eq. (S2) is closely related to the physics of optically bistable systems. In particular, removing the $d$-resonance, one has the canonical model of an optically bistable resonance (\cite{haus1984waves}).
    \item More broadly, the Hamiltonian of Eq. (S2) is quite generic, and should apply to open nonlinear systems beyond those considered here. For example, in superconducting qubit systems, nonlinearities can be remarkably high, and there are a great many experimental possibilities for reservoir engineering. Such systems may yield compelling platforms to realize the Fock- and sub-Poissonian state-generation effects discussed here.
    \item All results have been provided in the limit $\kappa \ll \gamma$, enabling adiabatic elimination. Generalization of our results beyond this regime is of clear fundamental interest. It is also highly relevant in cases for which the frequency sharpness of the end-mirror becomes sharper than the response time of the cavity mode.
    \item As we showed, extremely strong noise reduction can also be obtained in systems with sharply nonlinear gain. The development of practical proposals of systems to realize a sharp nonlinear gain is then of interest as a ``competing'' platform to realize the Fock- and sub-Poissonian state-generation effects here.
    \item We have focused almost exclusively on the quantum statistics of the cavity mode. The statistics of the output beam are also of obvious interest, and are simpler to probe than the cavity statistics. A detailed theoretical exposition of the output field statistics is therefore motivated.
    \item The entirety of the manuscript assumes that only a single-mode of the electromagnetic field is relevant. Treatments of Fock-state generation (with or without gain) in the multimode regime are of obvious interest.
\end{enumerate}

\clearpage

\section{Supplementary figures}

Here, we provide additional figures and results, as well as a table of detailed parameters used in Figs. 3, 4 of the main text.


\begin{figure*}[h]
    \centering
    \includegraphics[width=0.75\textwidth]{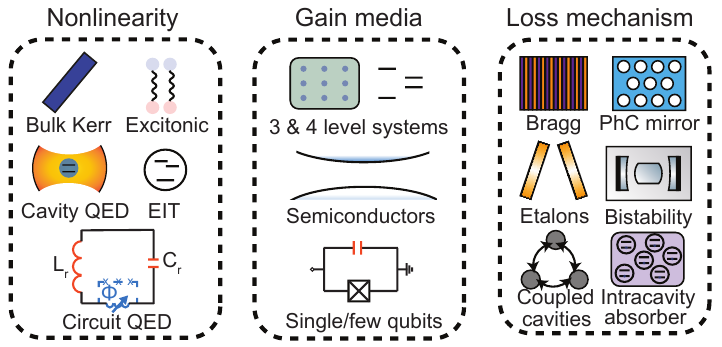}
    \caption{\textbf{Systems which could be explored for Fock lasing.} Many options exist for nonlinearity: circuit and cavity QED systems, atomic gases, excitonic strong coupling, and bulk optical materials. Gain media span solid-state, semiconductors, gases, dyes, artificial atoms, and even single atoms (in one-atom lasers). Sources of sharp loss include absorbers, as well as many systems explored in (nano)photonics: photonic crystals, Fano resonances, bound states in the continuum, bistable systems, and coupled cavities. }
\end{figure*}

In Table S1, we provide detailed parameters for the gain for the examples of Figs. 3, 4 of the main text. The various parameters to be specified are: the gain medium frequency $(\omega_{\text{gain}})$, relaxation and decoherence rates $\gamma_{||},\gamma_{\perp}$, the cross sections for stimulated emission and absorption ($\sigma_{\text{st}}, \sigma_{\text{abs}}$), the density of gain atoms $n_{\text{gain}}$, and the fill fraction $f$ of the gain. The cavity is specified by the cavity length $L_{\text{cav}}$, the cavity waist $w_{\text{cav}}$, the resonance frequency $\omega_{\text{cav}}$, and the nonlinear strength per photon $\beta$. The lasing mode is taken as a $\text{TEM}_{00}$ mode. The Fano mirror is parameterized by its width $\gamma$, its direct transmission coefficient $t_d$ (see Eq. (S38)), and its frequency $\omega_d = (1+\delta)\omega_a$.

\begin{table} [t]
\begin{tabular}{||c | c | c ||} 
 \hline
 Parameter & Value (Fig. 3) & Value (Fig. 4) \\ [0.5ex] 
 \hline\hline
 $\omega_{\text{gain}}$ & 1.47 eV & 1.17 eV  \\ 
 \hline
 $\gamma_{||}$ & $3 \times 10^{8}$ s$^{-1}$  & $4.34 \times 10^{3}$ s$^{-1}$  \\
 \hline
 $\gamma_{\perp}$ & $3.1 \times 10^{13}$ s$^{-1}$  & $1 \times 10^{12}$ s$^{-1}$ \\
 \hline
 $\sigma_{\text{st}}$ & $3 \times 10^{-16}$ cm$^{2}$ & $2.8 \times 10^{-19}$ cm$^{2}$  \\
 \hline
 $\sigma_{\text{abs}}$ & $3 \times 10^{-16}$ cm$^{2}$ & $7.7 \times 10^{-20}$ cm$^{2}$ \\
 \hline
 $n_{\text{gain}}$ & $1.7 \times 10^{21}$ cm$^{-3}$ & $1.3 \times 10^{20}$ cm$^{-3}$  \\
 \hline
  $f$ & 0.5 & 0.5  \\
 \hline
  $L_{\text{cav}}$ & 2 $\mu$m & 1 mm \\
 \hline
  $w_{\text{cav}}$ & 1 $\mu$m & 40 $\mu$m \\
 \hline
  $\omega_{\text{cav}}$ & $\omega_{\text{gain}}$ & $\omega_{\text{gain}}$ \\
 \hline
  $\beta$ & $-10^{-5}\omega_{\text{cav}}$ & $5 \times 10^{-18}\omega_{\text{cav}}$ \\
 \hline
  $\gamma$ & $2 \times 10^{-3}\omega_{\text{cav}}$ & $10^{-2}\omega_{\text{cav}}$ \\
 \hline
  $t_d$ & 0.05 & 1 \\
 \hline
 $\delta$ & $20\gamma$ & $-10^{-3}\gamma$  \\ [1ex] 
 \hline
\end{tabular}
\\
\caption{Table of gain, cavity, and linear resonance parameters used in Figs. 3, 4 of the main text.}
\end{table}

\begin{figure*}[h]
    \centering
    \includegraphics[width=1\textwidth]{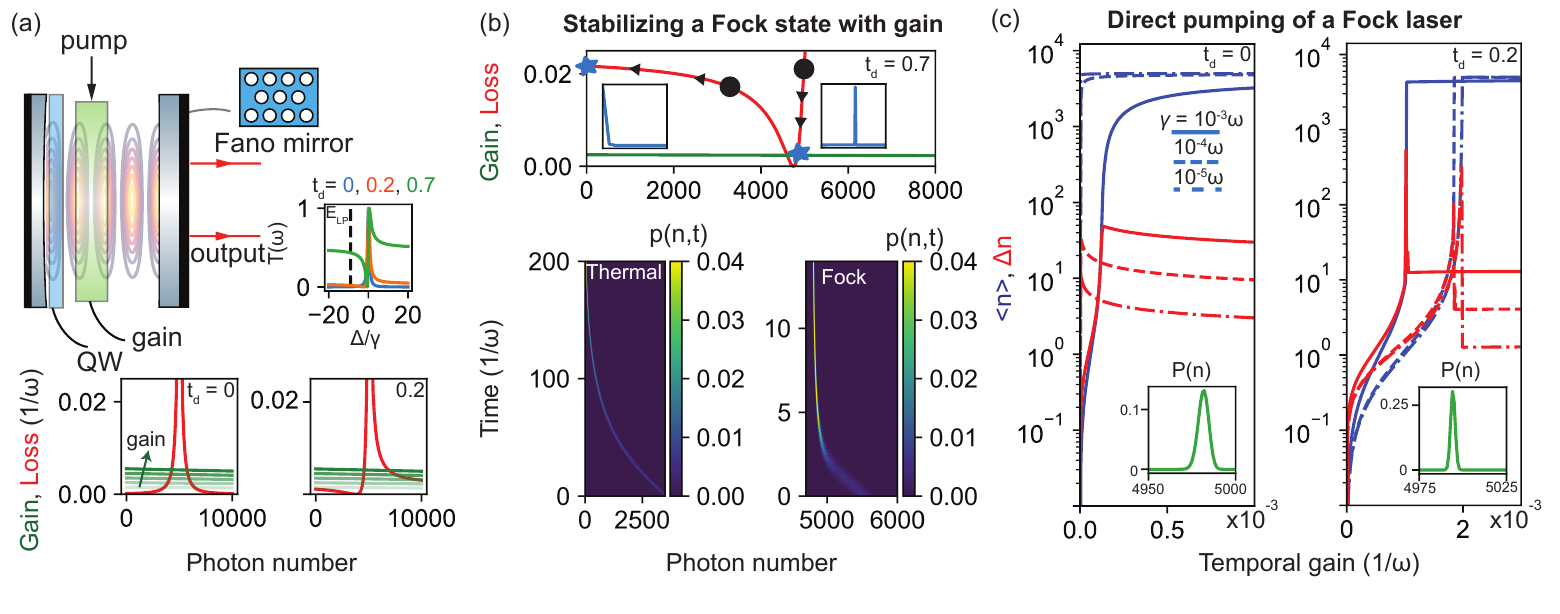}
    \caption{\textbf{Fock lasing in systems with strong optical nonlinearities.} (a) The system of Fig. 3 of the main text is now converted into a ``Fock laser'' by inclusion of a gain medium. Different transmission profiles for the Fano mirror lead to different losses, and thus different emission-absorption diagrams. Note that $t_d$ is the direct transmission coefficient that controls the Fano lineshape. (b) Evolution of an initial coherent state with different photon numbers (black circles) in the Fock laser. A state to the left of the approximate zero of the loss decays into a thermal state with a very low number of photons, while a state to the right of the zero decays into a steady-state with very low noise, approaching a high-number optical Fock state. (c) Photon number and fluctuations as a function of pump. ``S-curves'' similar to conventional lasers are observed in the photon number, except they saturate much more strongly, with the photon number hardly changing for increasing pump. Moreover, the photon number fluctuations, rather than increasing according to shot noise, decrease to nearly zero beyond threshold, indicating convergence to a near-Fock state. Different curves indicate different values of the mirror sharpness $\gamma$. In this figure, the polaritons have Kerr nonlinear strength $10^{-7}\omega_{\text{LP}}$. The detuning of the mirror from the lower polariton energy (with zero polaritons) is $10^{-3}\omega_0$ and the mirror has a sharpness of $10^{-4}\omega_0$. }
\end{figure*}
\clearpage

\section{Appendix: Deriving the effect of nonlinear loss on probabilities and coherences directly from the Heisenberg picture}

In Section II (``Quantum theory of a nonlinear resonator with frequency-dependent loss''), we derived the equation of motion for the photon probabilities from a reservoir theory in which we considered the joint coupling of the cavity and end-mirror to the resonator. We derived a master equation for the density matrix of the cavity and mirror and we then adiabatically eliminated the end mirror. We now provide a potentially simpler and more direct derivation of the result from the Heisenberg equations of motion. This derivation fully agrees with our findings from the density matrix. 

\subsection{General framework}

Our goal will be to derive a set of Heisenberg equations of motion to describe the photon in the nonlinear cavity. In a conventional laser theory based on Langevin equations, one writes an equation of motion for $a$. For the nonlinear laser considered here, this approach is complicated by the polychromatic nature of a nonlinear oscillator. In particular, the operator $a$ can be expressed as $a = \sum\limits_n \sqrt{n} |n-1\rangle\langle n| \equiv \sum\limits_n \sqrt{n} T_{n-1,n} $. In the absence of interactions with gain or reservoirs, the time-evolution of $a$ would simply be $a(t) = \sum\limits_n \sqrt{n} T_{n-1,n}(0)e^{-i\omega_{n,n-1}t}$ with $\omega_{n,n-1} = \omega_n - \omega_{n-1}.$ For a linear photon, $\omega_{n,n-1} =  n\omega - (n-1)\omega = \omega$, independently of $n$, recovering the familiar monochromatic evolution $a(t) = a(0)e^{-i\omega t}$. 


While the polychromatic nature of $a$ evades solution by conventional methods, the time evolution of the operators $T_{n-1,n}$, and more generally, $T_{n-k,n}$, is quite simple. For example, in the absence of gain or loss, the time-evolution of the operator $T_{n-k,n}$ is given as:
\begin{equation}
\dot{T}_{n-k,n} = \frac{i}{\hbar}\left[\sum\limits_m  \hbar\omega_m T_{m,m}, T_{n-k,n}\right] = -i\omega_{n,n-k}T_{n-k,n},
\end{equation}
so that $T_{n-k,n}(t) = T_{n-k,n}(0)e^{-i\omega_{n,n-k}t}$. Thus, the operators $T_{n-k,n}$ have a simple monochromatic evolution in the absence of interactions. The simplicity of the equation of motion for the projectors then motivates us to formulate our quantum theory of nonlinear loss through the equations of motion for the $T_{n-k,n}$, for each $k$. Each $k$ corresponds to a quantity with clear physical significance. The case of $k=0$, which is of primary interest in this work, corresponds to probabilities/populations. In particular, $\langle T_{n,n}\rangle = \text{tr}[\rho T_{n,n}]$ corresponds to the probability of having $n$ photons. The case of $k$ finite correspond to coherences, with $\langle T_{n-1,n}\rangle$ corresponding to first-order (phase) coherence (and the laser linewidth) and $\langle T_{n-2,n}\rangle$ corresponding to second-order (intensity) coherence. 

In deriving Eq. (S81), we have made use of the fundamental identity of projectors $T_{ij}T_{kl} = \delta_{jk}T_{il}$. We will make heavy use of this identity throughout this section. Beyond this, the following two identities are also used frequently:
\begin{align}
&\left[a, T_{n-k,n}\right] = \sqrt{n-k}T_{n-k-1,n} - \sqrt{n+1}T_{n-k,n+1},\\
&\left[a^{\dagger}, T_{n-k,n}\right] = \sqrt{n-k+1}T_{n-k+1,n} - \sqrt{n}T_{n-k,n-1}.
\end{align}

We have already found the contribution of free evolution to the equation of motion for $T_{n-k,n}$. Now we move to find the contribution from the sharp loss provided by the end mirror.

\subsection{Loss terms}

Now, we derive the contribution of cavity losses to the equation of motion for the $k$th coherences: defined as $\dot{T}^{(\text{loss})}_{n-k,n}$. We have
\begin{align}
\dot{T}^{(\text{loss})}_{n-k,n} &= i\left[(\lambda ad^{\dagger} + \lambda^* a^{\dagger}d) + \sum\limits_k g_k(ab_k^{\dagger} + a^{\dagger}b_k), T_{n-k,n}\right] \nonumber \\ 
&= i\sum_k g_k b_k^{\dagger}(\sqrt{n-k}T_{n-k-1,n} - \sqrt{n+1}T_{n-k,n+1})  \nonumber \\
&+ i\sum_k g_k (\sqrt{n-k+1}T_{n-k+1,n} - \sqrt{n}T_{n-k,n-1})b_k \nonumber \\
&+ i\lambda d^{\dagger}(\sqrt{n-k}T_{n-k-1,n} - \sqrt{n+1}T_{n-k,n+1}) \nonumber \\ 
&+ i\lambda^* (\sqrt{n-k+1}T_{n-k+1,n} - \sqrt{n}T_{n-k,n-1})d \nonumber \\
& \equiv (\text{L1A}) + (\text{L1B}) + (\text{L2A}) + (\text{L2B}).
\end{align}
Here, we have normally ordered the reservoir operators, as we will exclusively consider initial conditions involving no excitations in the far-field or the internal mode of the Fano mirror. Therefore, upon taking expectation values, terms involving the initial values of these operators (Langevin forces) will vanish. 

Now, we eliminate the reservoirs from the equations. This is done through the Heisenberg equations of motion for the far-field reservoir and the internal mode of the Fano mirror. The equation for $b_k$ reads:
\begin{equation}
\dot{b}_k = -i\omega_k b_k - ig_k a - iv_k d,
\end{equation}
admitting the formal solution
\begin{equation}
b_k(t) = b_k(0)e^{-i\omega_k t} - i\int\limits^t dt' \left(g_k a(t') + v_k d(t') \right)e^{-i\omega_k (t-t')}.
\end{equation}
To proceed, let us eliminate $b$ from the equation of motion for $d$. The equation of motion for $d$ is:
\begin{equation}
\dot{d} = -i\omega_d d - i\lambda a - i\sum\limits_k v_k b_k. 
\end{equation}
Plugging in the formal solution for $b_k$ results in:
\begin{equation}
\dot{d} = -i\omega_d d - i\lambda a - i\sum\limits_k v_k \left(b_k(0)e^{-i\omega_k t} - i\int\limits^t dt' \left(g_k a(t') + v_k d(t') \right)e^{-i\omega_k (t-t')} \right).
\end{equation}
Now, we make use of the fact that in laser theory, the coupling between cavity modes and the far-field is well-approximated as a white noise coupling which is independent of frequency, so that $g_k = g$ and $v_k = v$ (Markov approximation). In that case, the sum over $k$ can be carried out. In the continuum limit, $\sum_k \rightarrow \int d\omega_k ~\rho_0$, with $\rho_0$ the (constant) density of (far-field) states, such that the sum yields:
\begin{equation}
\dot{d} = -is_d d - G_-a + F_d.
\end{equation}
Here, we have used $\int dt' \delta(t-t')f(t') = \frac{1}{2}f(t)$ and defined $\gamma = 2\pi\rho v^2$,  $\kappa = 2\pi\rho g^2$, $s_d = \omega_d - i\frac{\gamma}{2}$, and $G_- = i\lambda + \frac{1}{2}\sqrt{\kappa\gamma}$. We have also defined the Langevin force on $d$ via $F_d = - i\sum\limits_k v_k b_k(0)e^{-i\omega_k t}$. We may now write the formal solution for $d$ as
\begin{equation}
d(t) = d(0)e^{-is_d t} + \int\limits^t dt' \left(-G_- a(t') + F_d(t') \right)e^{-is_d(t-t')}.  
\end{equation}
With the formal solutions for $b$ and $d$, we may now plug them back into the terms L1A, L1B, L2A, and L2B. Let us start with L1A and L1B. L1A , under the Markov approximation, is given as:
\begin{equation}
\text{(L1A)} = \left(i\sum_k g_k b^{\dagger}_k(0)e^{i\omega_k t} - \frac{1}{2}  \left(\kappa a^{\dagger} + \sqrt{\kappa\gamma} d^{\dagger} \right) \right)(\sqrt{n-k}T_{n-k-1,n} - \sqrt{n+1}T_{n-k,n+1}).
\end{equation}
To proceed, we carry out the following steps (these will be repeated for the terms L1B, L2A, and L2B): 
\begin{align}
\text{(L1A)} &= \left(i\sum_k g_k b^{\dagger}_k(0)e^{i\omega_k t}\right) (\sqrt{n-k}T_{n-k-1,n} - \sqrt{n+1}T_{n-k,n+1}) \nonumber \\
& - \frac{1}{2}  \kappa ((n-k)T_{n-k,n} - \sqrt{(n+1)(n-k+1)}T_{n-k+1,n+1})  \nonumber \\
& + \frac{1}{2} \sqrt{\kappa\gamma} \int\limits^t dt' G^*_- a^{\dagger}(t')e^{is^*_d(t-t')}(\sqrt{n-k}T_{n-k-1,n} - \sqrt{n+1}T_{n-k,n+1}) \nonumber \\
& - \frac{1}{2} \sqrt{\kappa\gamma} \left(d^{\dagger}(0)e^{is^*_d t} + \int\limits^t dt' F^{\dagger}_d(t') e^{is^*_d(t-t')}\right) (\sqrt{n-k}T_{n-k-1,n} - \sqrt{n+1}T_{n-k,n+1}).
\end{align}
In what follows, we consider the limiting case in which the decay of $d$, set by $\gamma$ is much faster than the gain dynamics. This is the same adiabatic approximation that was used in the density matrix treatment of the nonlinear Fano resonance. Under those conditions, the third term becomes:
\begin{equation}
\frac{1}{2} \sqrt{\kappa\gamma} G^*_- \left(\frac{n-k}{i(\omega_{n-k,n-k-1}-s_d^*)}T_{n-k,n} - \frac{\sqrt{(n+1)(n-k+1)}}{i(\omega_{n-k+1,n-k}-s_d^*)}T_{n-k+1,n+1}\right).
\end{equation}
This allows us to write L1A as
\begin{align}
\text{(L1A)} = &-\frac{1}{2}  \kappa ((n-k)T_{n-k,n} - \sqrt{(n+1)(n-k+1)}T_{n-k+1,n+1})  \nonumber \\
& + \frac{1}{2} \sqrt{\kappa\gamma} G^*_- \left(\frac{n-k}{i(\omega_{n-k,n-k-1}-s_d^*)}T_{n-k,n} - \frac{\sqrt{(n+1)(n-k+1)}}{i(\omega_{n-k+1,n-k}-s_d^*)}T_{n-k+1,n+1}\right). \nonumber \\
& +\left(i\sum_k g_k b^{\dagger}_k(0)e^{i\omega_k t}\right) (\sqrt{n-k}T_{n-k-1,n} - \sqrt{n+1}T_{n-k,n+1}) \nonumber \\
& - \frac{1}{2} \sqrt{\kappa\gamma} \left(d^{\dagger}(0)e^{is^*_d t} + \int\limits^t dt' F^{\dagger}_d(t') e^{is^*_d(t-t')}\right) (\sqrt{n-k}T_{n-k-1,n} - \sqrt{n+1}T_{n-k,n+1}).
\end{align}
As can be seen, the first two lines, upon taking expectation values, give terms of a similar form to those derived for the density matrix. The remaining lines give zero expectation value when starting in the vacuum of the internal mode and the reservoir, and thus vanish when considering equations of motion for coherences. 

Now, let us consider the remaining terms. L1B is quite similar to L1A, and we write 
\begin{equation}
\text{(L1B)} = (\sqrt{n-k+1}T_{n-k+1,n} - \sqrt{n}T_{n-k,n-1})\left(i\sum_k g_k b_k(0)e^{-i\omega_k t} + \frac{1}{2}  \left(\kappa a + \sqrt{\kappa\gamma} d \right) \right),
\end{equation}
which may be further simplified as
\begin{align}
\text{(L1B)} &= \frac{1}{2}\kappa (\sqrt{(n-k+1)(n+1)}T_{n-k+1,n+1} - nT_{n-k,n}) \nonumber \\
&- \frac{1}{2}\sqrt{\kappa\gamma}G_-\left(\frac{\sqrt{(n-k+1)(n+1)}}{i(s_d-\omega_{n+1,n})}T_{n-k+1,n+1} - \frac{n}{i(s_d-\omega_{n,n-1} )}T_{n-k,n}\right) \nonumber \\
&+ (\sqrt{n-k+1}T_{n-k+1,n} - \sqrt{n}T_{n-k,n-1})\left(i\sum_k g_k b_k(0)e^{-i\omega_k t} \right) \nonumber \\
&+ \frac{1}{2}\sqrt{\kappa\gamma}(\sqrt{n-k+1}T_{n-k+1,n} - \sqrt{n}T_{n-k,n-1})\left(d(0)e^{-is_d t} + \int\limits^t dt' F_d(t') e^{-is_d(t-t')} \right),
\end{align}
where we have taken all the same steps as those leading to Eq. (S94). 

The term L2A is given as:
\begin{equation}
\text{(L2A)} = i\lambda \left(d^{\dagger}(0)e^{is^*_d t} + \int\limits^t dt' \left(-G^*_- a^{\dagger}(t') + F^{\dagger}_d(t') \right)e^{+is^*_d(t-t')} \right)(\sqrt{n-k}T_{n-k-1,n} - \sqrt{n+1}T_{n-k,n+1}).
\end{equation}
Under the adiabatic approximation, we may then write:
\begin{align}
\text{(L2A)} &= -i\lambda G^*_-  \left(\frac{(n-k)}{i(\omega_{n-k,n-k-1}-s_d^*)}T_{n-k,n} - \frac{\sqrt{(n-k+1)(n+1)}}{i(\omega_{n-k+1,n-k}-s_d^*)}T_{n-k+1,n+1}\right) \nonumber \\
&+ i\lambda\left(d^{\dagger}(0)e^{is^*_d t} + \int\limits^t dt' F^{\dagger}_d(t') e^{+is^*_d(t-t')} \right)(\sqrt{n-k}T_{n-k-1,n} - \sqrt{n+1}T_{n-k,n+1}).
\end{align}

The term L2B:
\begin{equation}
\text{(L2B)} = i\lambda^* (\sqrt{n-k+1}T_{n-k+1,n} - \sqrt{n}T_{n-k,n-1})\left(d(0)e^{-is_d t} + \int\limits^t dt' \left(-G_- a(t') + F_d(t') \right)e^{-is_d(t-t')} \right),
\end{equation}
similarly follows as:
\begin{align}
\text{(L2B)} &= -i\lambda^* G_- \left(\frac{\sqrt{(n-k+1)(n+1)}}{i(s_d-\omega_{n+1,n})}T_{n-k+1,n+1} - \frac{n}{i(s_d-\omega_{n,n-1})}T_{n-k,n} \right)   \nonumber \\
&+ i\lambda^* (\sqrt{n-k+1}T_{n-k+1,n} - \sqrt{n}T_{n-k,n-1})\left(d(0)e^{-is_d t} + \int\limits^t dt'  F_d(t') e^{-is_d(t-t')} \right).
\end{align}

Plugging L1A, L1B, L2A, and L2B into the equation for $\dot{T}^{(\text{loss})}_{n-k,n}$, we have
\begin{align}
\dot{T}^{(\text{loss})}_{n-k,n} &= \left(-\frac{1}{2}\kappa (2n-k) +  \frac{(n-k)\left(-i\lambda + \frac{1}{2} \sqrt{\kappa\gamma}\right) G^*_- }{i(\omega_{n-k,n-k-1}-s_d^*)}  - \frac{n \left(-i\lambda^* - \frac{1}{2} \sqrt{\kappa\gamma}\right)G_-}{i(s_d-\omega_{n,n-1} )} \right)T_{n-k,n} \nonumber \\
&+ \sqrt{(n-k+1)(n+1)}\left(\kappa  - \frac{\left(-i\lambda + \frac{1}{2} \sqrt{\kappa\gamma}\right) G^*_-}{i(\omega_{n-k+1,n-k}-s_d^*)} + \frac{\left(-i\lambda^* - \frac{1}{2} \sqrt{\kappa\gamma}\right)G_-}{i(s_d-\omega_{n+1,n})} \right)T_{n-k+1,n+1} \nonumber \\
& + F^{(\text{loss})}_{n-k,n},
\end{align}

The Langevin force $F^{(\text{loss})}_{n-k,n}$ is given by
\begin{align}
F^{(\text{loss})}_{n-k,n} &= \left(i\sum_k g_k b^{\dagger}_k(0)e^{i\omega_k t}\right) (\sqrt{n-k}T_{n-k-1,n} - \sqrt{n+1}T_{n-k,n+1}) \nonumber \\
& + (\sqrt{n-k+1}T_{n-k+1,n} - \sqrt{n}T_{n-k,n-1})\left(i\sum_k g_k b_k(0)e^{-i\omega_k t} \right) \nonumber \\
& +  \left(i\lambda - \frac{1}{2} \sqrt{\kappa\gamma}\right) \left(d^{\dagger}(0)e^{is^*_d t} + \int\limits^t dt' F^{\dagger}_d(t') e^{is^*_d(t-t')}\right) (\sqrt{n-k}T_{n-k-1,n} - \sqrt{n+1}T_{n-k,n+1}) \nonumber \\
& + \left(i\lambda^* + \frac{1}{2}\sqrt{\kappa\gamma}\right)(\sqrt{n-k+1}T_{n-k+1,n} - \sqrt{n}T_{n-k,n-1})\left(d(0)e^{-is_d t} + \int\limits^t dt' F_d(t') e^{-is_d(t-t')} \right),
\end{align}
and has the important property that $\langle F^{(\text{loss})}_{n-k,n} \rangle = 0$ when the initial state is the vacuum of the reservoirs and the internal mode. Hence, for the systems we will consider here, such terms can be functionally ignored.


\subsection{Equation of motion for the $k$-th coherences}

Here, we summarize the previous two sections, writing down the total equations of motion for the photon. The equation of motion for the $k$th coherences are
\begin{align}
\dot{T}_{n-k,n} &= -i\omega_{n,n-k}T_{n-k,n} \nonumber \\
&+  \left(-\frac{1}{2}\kappa (2n-k) +  \frac{(n-k)(G_+G_-)^* }{i(\omega_{n-k,n-k-1}-s_d^*)}  + \frac{n G_+G_-}{i(s_d-\omega_{n,n-1} )} \right)T_{n-k,n} \nonumber \\
&+ \sqrt{(n-k+1)(n+1)}\left(\kappa  - \frac{(G_+G_-)^*}{i(\omega_{n-k+1,n-k}-s_d^*)} + \frac{G_+G_-}{i(s_d-\omega_{n+1,n})} \right)T_{n-k+1,n+1} \nonumber \\
& + F^{(\text{loss})}_{n-k,n},
\end{align}
where
\begin{align}
F^{(\text{loss})}_{n-k,n} &= \left(i\sum_k g_k b^{\dagger}_k(0)e^{i\omega_k t}\right) (\sqrt{n-k}T_{n-k-1,n} - \sqrt{n+1}T_{n-k,n+1}) \nonumber \\
& + (\sqrt{n-k+1}T_{n-k+1,n} - \sqrt{n}T_{n-k,n-1})\left(i\sum_k g_k b_k(0)e^{-i\omega_k t} \right) \nonumber \\
& +  \left(i\lambda - \frac{1}{2} \sqrt{\kappa\gamma}\right) \left(d^{\dagger}(0)e^{is^*_d t} + \int\limits^t dt' F^{\dagger}_d(t') e^{is^*_d(t-t')}\right) (\sqrt{n-k}T_{n-k-1,n} - \sqrt{n+1}T_{n-k,n+1}) \nonumber \\
& + \left(i\lambda^* + \frac{1}{2}\sqrt{\kappa\gamma}\right)(\sqrt{n-k+1}T_{n-k+1,n} - \sqrt{n}T_{n-k,n-1})\left(d(0)e^{-is_d t} + \int\limits^t dt' F_d(t') e^{-is_d(t-t')} \right).
\end{align}
One immediately sees that for $k=0$, these equations are identical to those from the density matrix description $-$ modulo the explicit form of the Langevin terms here, which resulted from our explicit account of the reservoir in the Heisenberg equations.

\bibliographystyle{unsrt}
\bibliography{Fock_laser.bib}

\end{document}